\documentclass[a4paper,UKenglish,cleveref,autoref,thm-restate]{tgdk-v2021}

\usepackage{graphicx}
\usepackage[colorinlistoftodos]{todonotes}

\usepackage{adjustbox}
\usepackage{caption}
\usepackage{multirow}
\usepackage{makecell}
\usepackage{subcaption}
\usepackage{soul,xcolor}

\title{Talking Wikidata: Communication patterns and their impact on community engagement in collaborative knowledge graphs} 

\titlerunning{Talking Wikidata} 

\author{Elisavet Koutsiana}{King's College London, Bush House, Strand, London}{elisavet.koutsiana@kcl.ac.uk}{https://orcid.org/0000-0001-6544-0435}{European Union’s Horizon 2020 research and innovation program under the Marie Skłodowska-Curie grant agreement no 812997 (CLEOPATRA ITN)}

\author{Ioannis Reklos}{King's College London, Bush House, Strand, London}{ioannis.reklos@kcl.ac.uk}{https://orcid.org/0000-0002-2747-579X}{}

\author{Kholoud {Saad Alghamdi}}{King's College London, Bush House, Strand, London}{kholoud.alghamdi@kcl.ac.uk}{https://orcid.org/0000-0003-0261-9977}{}

\author{Nitisha Jain}{King's College London, Bush House, Strand, London}{nitisha.jain@kcl.ac.uk}{https://orcid.org/0000-0002-7429-7949}{}

\author{Albert Meroño-Peñuela}{King's College London, Bush House, Strand, London}{albert.merono@kcl.ac.uk}{https://orcid.org/0000-0003-4646-5842}{}

\author{Elena Simperl}{King's College London, Bush House, Strand, London}{elena.simperl@kcl.ac.uk}{https://orcid.org/0000-0003-1722-947X}{}

\authorrunning{Koutsiana et al.} 

\Copyright{Elisavet Koutsiana, Ioannis Reklos, Kholoud S. Alghamdi, Nitisha Jain, Alabert M. Peñuela, Elena Simperl} 



\begin{document}
\nolinenumbers
\setstcolor{red}
\maketitle

\begin{abstract}
We study collaboration patterns of Wikidata, one of the world's largest open source collaborative knowledge graph (KG) communities. 
Collaborative KG communities, play a key role in structuring machine-readable knowledge to support AI systems like conversational agents. However, these communities face challenges related to long-term member engagement, as a small subset of contributors often is responsible for the majority of contributions and decision-making. While prior research has explored contributors' roles and lifespans, discussions within collaborative KG communities remain understudied. To fill this gap, we investigated the behavioural patterns of contributors and factors affecting their communication and participation. We analysed all the discussions on Wikidata using a mixed methods approach, including statistical tests, network analysis, and text and graph embedding representations. Our findings reveal that the interactions between Wikidata editors form a small world network, resilient to dropouts and inclusive, where both the network topology and discussion content influence the continuity of conversations. Furthermore, the account age of Wikidata members and their conversations are significant factors in their long-term engagement with the project. 
Our observations and recommendations can benefit the Wikidata and semantic web communities, providing guidance on how to improve collaborative environments for sustainability, growth, and quality.

\end{abstract}

\section{Introduction}

Online peer-production communities are powerful tools with groups of self-organised people who work collaboratively towards a shared outcome. They offer innovation, efficiency, knowledge sharing, and democratisation of technology. Over the years, communities like those of Quora,\footnote{\url{https://www.quora.com/}} Stack Overflow,\footnote{\url{https://stackoverflow.com/}} Wikipedia,\footnote{\url{https://www.wikipedia.org/}} and Wikidata,\footnote{\url{https://www.wikidata.org/wiki/Wikidata:Main_Page}} have made significant contributions to computer science. 
Collaborative knowledge graph (KG) communities, such as Wikidata, are particularly important as they provide structured knowledge from large datasets \cite{zhou2020survey} in machine-readable way to support various AI systems, including conversational agents and voice assistants \cite{fensel2020knowledge}.
The community plays a significant role in the maintenance, sustainability, and success of collaborative KGs \cite{benkler2015peer}. However, these communities often face the challenge of long-term engagement from members \cite{sarasua2019evolution,piscopo2018models}, with a small group responsible for the majority of contributions and decision-making \cite{muller2015peer,piao2021learning}. Therefore, it is essential to study the collaborative mechanisms behind contributors to understand what impacts their participation and contributions.

A plethora of online discussion studies in peer-production communities showed that contributors' discussions are a valuable source of insights \cite{viegas2007talk,hata2022github,moutidis2021community}. At the same time, the variety of types and editing processes of peer-production systems, such as an online encyclopedia, an open-source software system, and a KG, have raised concerns about the applicability of assumptions across different projects \cite{filippova2016effects,kittur2010beyond,wikimediaContoversies}. As a result, it is essential to investigate the different types of peer-production systems individually to better understand their unique characteristics and challenges.
For collaborative KGs, analysis of discussions is missing from the line of research, with previous work focusing primarily on emerging roles \cite{muller2015peer,piscopo2018models} and lifespan \cite{sarasua2019evolution} based on their editing activities and semi-structured interviews.

To fill this gap, we investigated members' behaviour in large-scale collaborative KGs.
Specifically, we studied the Wikidata community and examined how members' communication patterns affected the project.
Wikidata is an open generic interest KG supporting numerous intelligent systems \cite{vrandecic2013rise}. It has a diverse community of volunteers who collaborate online. 
We analysed the behaviour of Wikidata contributors by answering:
\begin{itemize}
    \item RQ1 - What are the characteristics of editors’ collaborations?
    \item RQ2 - What factors affect whether a discussion will receive a response?
    \item RQ3 - Do discussions affect editor engagement?
\end{itemize}

To understand community interactions, we used a mixed-method approach with descriptive statistics, statistical tests, network analysis, as well as neural networks that combined graph and text embeddings. Previous work in online communities, to understand the characteristics of contributors' collaborations, analysed their communication using network analysis \cite{leskovec2008planetary,ingawale2009small,fisher2017perceived}, and their behaviour patterns through members characteristics, such as volume of edits, posts, or account age \cite{brandes2009network,kanza2018does}. We followed similar practices to answer $RQ1$. Additionally, to answer $RQ2$ and $RQ3$ and identify the factors affecting communication and participation we develop a machine learning model combining features of members' characteristics, their posts using text embedding vectors and their topology using graph embedding vectors.

Our contributions are threefold: 
\begin{itemize}
    \item we present one of the first discussion studies on the interaction of collaborative KG communities
    \item we publish a dataset of posts of the Wikidata discussions
    \item we propose a framework to investigate what influences conversations and editors' engagement
    \item we recommend a list of practices to improve communication and participation
\end{itemize}

Our study indicates that discussion interactions between Wikidata editors form a network with a high clustering coefficient and low shortest path, suggesting a small world network. 
In this network, the network topology and the content of discussions influence the continuation of conversations. At the same time, the age of an editor in the community and their discussions influence their engagement. Our findings can support Wikidata in improving its practices and tools. Based on our findings, we recommend the development of new systems to track posts and assist in writing/asking about complex topics like KE. We also highlight the importance of improvements for assisting new members and enhancing engagement.
Additionally, our results can aid recommendation systems, such as those found in \cite{zangerle2016empirical,alghamdi2021learning}, in design, for example, by using advice from editors found to be long-lasting in the community and in disseminating new tools with the assistance of well connected and popular members.

\section{Background: The Wikidata knowledge construction project}
\label{sec:background}

\subsection{The knowledge graph} 
The term KG describes a knowledge base that is built and organised as a graph \cite{hogan2021knowledge} consisting of nodes (entities like \textit{people} and \textit{cities}) and edges (relations like \textit{friend of} or \textit{capital of}).
Wikidata is a free and openly available KG that people can use and contribute to \cite{vrandevcic2014wikidata}. It is part of the Wikimedia Foundation\footnote{\url{https://wikimediafoundation.org/about/}} and supports other projects like Wikipedia\footnote{\url{https://www.wikipedia.org/}}, Wikisource\footnote{\url{https://en.wikisource.org/wiki/Main_Page}}, etc. Today, Wikidata has more than $100M$ nodes called \textit{items}, $10K$ edge labels called \textit{properties}, and a community of more than $24K$ contributors, referred to as \textit{editors}.\footnote{\url{https://stats.wikimedia.org/\#/metrics/wikidata.org}} Items and properties are organised in taxonomies using the properties \textit{instance of} and \textit{subclass of}, similar to \textit{rdf:type} and \textit{rdfs:subClassOf}, respectively.

\subsection{The contributors} 
Editors can be either \textit{humans} or \textit{bots}. Humans can be registered community members or may choose to contribute anonymously. Registered members can gain additional editing rights, based on their experience with the project.\footnote{\url{https://www.wikidata.org/wiki/Wikidata:User\_access\_levels}}
Table \ref{tab:access_levels} presents the additional editing rights, called \textit{access levels}, found for editors participating in discussions.
For example, \textit{Administrators} access level is a group of editors with high administrative and technical rights elected by a community vote. Among others, they can grant rights to others, block spammers and prevent vandalism. 

Bots are software programs responsible for repetitive tasks like modification of items, or property constraint violations. Their services require approval from the community, and Wikidata currently has a total of $399$ bots, with approximately $100$ being active every month.\footnote{\url{https://stats.wikimedia.org/\#/metrics/wikidata.org}}

\begin{table}[h]
\scriptsize
\caption{List of access levels and their descriptions (\href{https://www.wikidata.org/wiki/Wikidata:User_access_levels}{source}).}
\label{tab:access_levels}
\begin{tabular}{|p{0.18\linewidth}|p{0.75\linewidth}|}
\hline
\textbf{Access level} &
  \textbf{Description} \\ \hline
Basic level users &
  Editors with basic editing rights. e.g., create and edit items. \\ \hline
Confirmed users &
  Editor accounts who have been confirmed as safe. An editor account with at least 4 days of age and at least 50 edits become autoconfirmed. Editors which do not meet the autoconfirmed criteria can be granted the confirmed manually. \\ \hline
Property creators &
  Editors who can create new properties. New properties can only be created by users in this group or administrators, after discussion in the respective communication page. \\ \hline
Rollbackers &
  Editors who can control the rollback tool. Rollback is a simple tool that can allow fast reverting of vandalism or other obvious abuse. \\ \hline
IP block exempt users &
  If an IP address is blocked with the option not to allow registered users to edit, this permission allows editing from such IP addresses \\ \hline
Administrators &
  These are trusted editors allowed to proceed with essential tasks such as blocking, or creating properties. They can also assign access rights to others, like rollback, confirmed user, IP block exempt, autopatrolled, and property creator. \\ \hline
Interface administrators &
  Editors with the ability to edit sitewide CSS/JS pages, change how content is styled, and change the behavior of pages. \\ \hline
Translator administrators &
  Editors who translate content to non English languages. Because Wikidata is a multilingual wiki, it uses the translate extension and translation administrators are users with access to administrative functions of the extension \\ \hline
Bureaucrats &
  They can grant or remove rights for administrators, bureaucrats, bots, flooders, and translator administrators. \\ \hline
CheckUsers &
  These editors examine the IP address data of registered users, which is otherwise private information. Its primary purpose is to investigate sock-puppets. \\ \hline
Suppressors &
  Suppression refers to hiding revisions, usernames in edit histories and logs, or portions of individual log entries. Suppressors can have access to these data. \\ \hline
Flooders &
  Users may be granted a "flood flag" and be flooders when doing repetitive, non-controversial edits. This is to distinguish those actions from vandalism. \\ \hline
Wikidata staff &
  To facilitate the development of Wikidata, some Wikimedia Deutschland staff members technically hold full administrator access. \\ \hline
\end{tabular}
\end{table}

\subsection{The contributions}
People contribute to the KG in various ways. They can edit by creating, revising, and removing content, or discuss by asking, answering, and deciding how to structure information. Wikidata is built using a wiki, which is a web platform that allows people to edit the project using a web page in the browser \cite{wagner2004wiki}. Each item and property within Wikidata has its own wiki page, which progresses over time.  
All the changes made to the wiki pages are publicly accessible. These changes are referred to as \textit{revisions} and they include metadata such as the name of the editor responsible for the revision, the page name, and a timestamp.

\subsection{The communication}
Similar to collaborative ontology engineering projects \cite{simperl2014collaborative}, Wikidata discussions are a tool for coordination and collaboration. Two types of discussion channels exist in Wikidata: \textit{talk pages} and \textit{communication pages}.\footnote{\url{https://www.wikidata.org/wiki/Wikidata:Community_portal}} 

Talk pages are available for every content page, such as item, property, or help page. Editors use talk pages to coordinate activities on the content page, report mistakes, and raise questions. In addition, various communication pages in Wikidata allow for discussion and decision-making regarding general concerns, technical or management support. Examples of communication pages include the \textit{Bot request} page, where people specify the need for bot development, or the \textit{Property proposal} and \textit{Properties for deletion} pages for graph construction needs. 

\begin{figure}[h]
\centering
\includegraphics[width=0.9\linewidth]{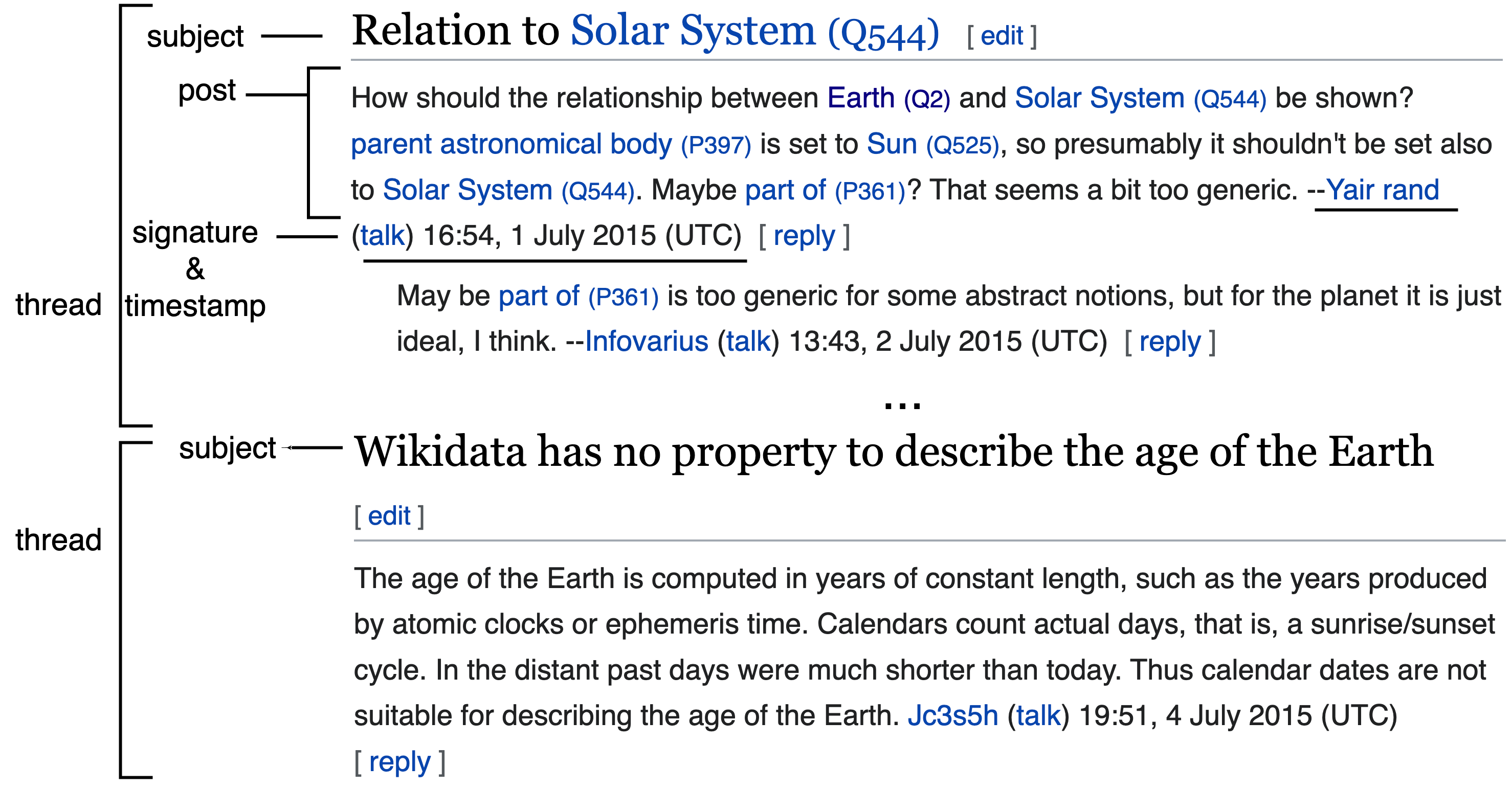}
\caption{An example of Wikidata discussions.} 
\label{fig:front_talk}
\end{figure}

Figure \ref{fig:front_talk} shows an example of a Wikidata discussion. Both the talk and communication pages follow a similar structure, with titles referred to as \textit{subjects} separating the raised issues and a sequence of responses referred to as \textit{posts}. Every post should end with the editors' username or IP, in the case of anonymous contribution, and a timestamp. A set of posts under a subject title is referred to as a \textit{thread}.

\section{Related work}

\subsection{Online communities through the lens of network analysis}

Previous studies on social media, online forums, and peer-production communities, has investigated their communication patterns. Several studies have focused on antisocial behaviour, such as anomaly detection (e.g., rumour, spamming, fraud, and cybercrime detection) \cite{ma2021comprehensive,pourhabibi2020fraud}, sentiment detection (e.g., irony or sarcasm, and controversy detection) \cite{joshi2017automatic,garimella2018quantifying}, and manipulation (e.g. sock-puppets) \cite{kumar2017army,solorio2013case} to maintain healthy communities. 
In addition, other studies analysed the structure of communication networks. 
For example, before the advent of social networks ($2006$), instant messaging platforms like MSN engaged $245M$ members in $30B$ conversations. An analysis of the MSN members' network \cite{leskovec2008planetary}, in which nodes represent members, and edges connect two members if they have ever exchanged messages, found characteristics like a power law in degree distribution; a high clustering coefficient, $0.1$; and one connected component including $>99\%$ of all nodes. These are similar to the characteristics of a small world network \cite{watts1998collective}, in which node distance grows proportionally to the logarithm of the number of nodes, rather than those of random networks \cite{erdos1959random}. 

Similar research was conducted examining the Slashdot site \cite{gomez2008statistical}. This is a news website for publishing. Readers can comment and score the posts. The authors built a network in which nodes represented members and edges connected to members if they responded to each other. The Slashdot network showed characteristics of a small world network, similar to the MSN network. The authors found a giant component containing $97\%$ of the nodes, a clustering coefficient of $0.05$ and an average shortest path of $4$ \cite{gomez2008statistical}. The study compared Slashdot with the characteristics of a typical social network using degree assortativity, which measured whether or not highly connected users were preferentially linked to other highly connected ones \cite{newman2003social}. Unlike social networks, which present strong assortativity with highly connected users connect with other highly connected users, Slashdot presented neutral assortativity. This means that members did not show any preference in responding based on the `popularity' of their peers. 
Degree assortativity seems to be a fundamental difference between social and non-social networks \cite{newman2003social}, as shown in similar studies on Facebook and Twitter \cite{fisher2017perceived,ugander2011anatomy}.

Prior work on Wikipedia, a community which shares similarities with Wikidata, found that Wikipedia exhibits characteristics of a small world network \cite{ingawale2009small}. In this analysis, nodes represented members and edges connected members who edited the same article. The network presented had a clustering coefficient of $0.9$ and an average shortest path of $5$ \cite{ingawale2009small}. Furthermore, the authors observed neutral assortativity which is similar to the Slashdot community. However, another study built communication networks for the Wikipedia Wikiprojects, which are small groups where contributors collaborate to build articles under a specific topic, and found that the networks were mostly disassortative, with `popular' editors communicating with less connected editors \cite{rychwalska2020quality}.
This may suggest that Wikiprojects have distinct characteristics from the other discussions, perhaps due to closely collaborating editors organising their work. This finding was similar to the distinction of topics discussed in Wikidata Wikiprojects \cite{kanke2020knowledge} and the item and property talk pages, and project chat communication pages \cite{koutsiana2023analysis}. The topics in Wikiprojects are likely to be related to organisation and management, while the talk pages are related to KE activities.
Additionally, an investigation of the temporal effect of assortativity in social networks, including Wikipedia, revealed that the assortativity increased rapidly at the early stages of the network growth and then decayed and remained stable after certain links were created \cite{zhou2020universal}.

Previous studies have explored the network structure in a variety of online communities. However, there is no prior work investigating the network dynamics of a collaborative KG. Many online communities have been shown to exhibit characteristics of small world networks, which has significant implications for how information is spread through the network. This includes efficient information sharing, exposure to diverse ideas, enhanced problem solving, faster epidemic control (vandalism), and resilience to editor drop outs  \cite{ingawale2009small,watts1998collective}. 
Additionally, assortativity has provided valuable insights into how people connect.
A similar investigation of the Wikidata community could reveal insight into information spread and group dynamics. 
This understanding could guide strategic decisions, improve engagement, enhance robustness (e.g., staying robust to random editor drop outs), and facilitate efficient information dissemination. For example, high assortativity is an indication of strong subgroups or cliques within the network. In this case, community managers might focus on engaging sub-communities separately. In contrast, low assortativity indicates an inclusive community where there is a flow of information between different levels of user engagement. In this case, broader community-wide engagement strategies could be more effective. These findings could support Wikidata in adopting better practices for effective communication and collaboration to produce high-quality outcomes and to encourage continued editor involvement.

\subsection{Member characteristics in peer-production communities}

Previous studies have explored behaviour patterns using member characteristics in peer-production communities to understand their underlying mechanisms of collaboration \cite{brandes2009network,kanza2018does}. In the case of Wikipedia, prior work studied interactions between users when editing articles by building edit networks to connect editors working on the same articles. \cite{brandes2009network}. The authors used characteristics like the type of the edit (e.g., delete, undelete, restore) and content (e.g., number of words added or deleted) to understand the work of individual editors and their controversial interactions.
For the community behind Schema.org, an ontology engineering project, prior work explored discussion participation patterns \cite{kanza2018does}. The authors created user profiles using characteristics like the number of emails sent and replied to or the issues raised. The study revealed, that  $10\%$ of users are responsible for $80\%$ of all contributions in community discussions. Additionally, the number of edits made was found to be positively correlated with the level of discussion activity \cite{falconer2011analysis}. This suggested that individuals who made many changes also participated in many discussions.

Previous studies in Wikidata used editing activities and user characteristics to analyse the behaviour of editors. By considering the number of edits, type of edits, type of editor (registered, anonymous, or bot), and account age, 
Muller et al. \cite{muller2015peer} found that most editors made specialised contributions, and only a small piece contributed widely to the project. Similar results were found by Piao et al. \cite{piao2021learning}, who showed that $0.08\%$ editors contributed $80\%$ of all edits. 
Sarasua et al. \cite{sarasua2019evolution} focused on the volume of edits for over time and argued that editors with high account age and high volumes of edits had the greatest impact on the project.
In addition, Piscopo et al. \cite{piscopo2018models} by combining editing activity with patterns in community discussions, revealed that editors presented diverse participation in editing and discussing. Our analysis combines participation levels with the duration of engagement, the editing rights of editors (Table \ref{tab:access_levels}) and discussion patterns to provide more insights into the behaviour of editors and engagement patterns. This understanding can be used to tailor engagement strategies

Previous studies used a variety of characteristics to investigate user behaviour in peer-production projects. However, several studies argued that due to the different editing processes of the projects, like an online encyclopedia, an open-source software system, and a KG, the projects present fundamental differences in user collaboration and interactions \cite{filippova2016effects,kittur2010beyond}. Particularly for Wikidata and Wikipedia, previous work highlighted the differences between editing an abstract article and editing a structured graph \cite{wikimediaContoversies,koutsiana2023agreeing}.

\subsection{The factors affecting members in peer-production systems}

Previous work has explored how discussions influence peer-production communities.
Studies in open source software communities \cite{filippova2016effects} and Wikipedia \cite{arazy2011information,arazy2013stay} have shown that different types of conflicts in discussions, such as task, normative, and process conflicts, affected the projects differently. Normative conflicts (i.e., group function conflicts) negatively impacted the intention to remain in the project, and process conflicts (i.e., disagreeing on how to perform tasks) harmed contributors' performance. Additionally, Arazy et al. \cite{arazy2013stay} argued that often, task conflicts transition to process conflicts, affecting the quality of Wikipedia articles. 
Finally, Rychwalska et al. \cite{rychwalska2020quality}  showed that the quality of Wikipedia articles was related to the private communication between editors.

Prior work has also explored the factors influencing editors to drop out of Wikidata.
Piao et al. \cite{piao2021learning} investigated engagement in Wikidata by studying the editing activities. The authors used various statistical and pattern-based features, such as the number of edits, activity types, duration between edits, pair of entities editing and their characteristics. They found that the feature that had the greatest impact on predicting dropout was the total number of edits made in the last few months. This is similar to findings in Wikipedia \cite{gandica2015wikipedia}. However, there is no prior evidence of how discussions influence editors' behaviour in Wikidata.



\section{Data}
\label{sec:data}

The Wikimedia Foundation supports publicly available dumps, updated monthly, for all its hosted projects. In our work, we used the XML dumps with the history of all edits.\footnote{\url{https://www.wikidata.org/wiki/Wikidata:Database_download}} For our work, we utilised the XML files that were updated until February 2023 with all activity logs and discussions. 

We used the XML files to generate two sets of data: a \textit{discussions} dataset that contained all posts in Wikidata and for each post we recorded the text, username, timestamp, and thread subject included; and a \textit{history of edits} dataset that included information for editors edits, referred to as revisions, and for each edit we recorded the username and timestamp (see Table \ref{tab:ds_editors_categories} for further descriptive statistics).

Our code and dataset are available on GitHub.\footnote{\url{https://github.com/ElisavetK/Talking_Wikidata_graph.git}}

\begin{table}[h]
\caption{Descriptive statistics for the editors' categories.}
\centering
\label{tab:ds_editors_categories}
\scriptsize
\begin{tabular}{|p{0.4\linewidth}|l|l|l|}
\hline
\textbf{Category} & \textbf{\#editors} & \textbf{\#posts} & \textbf{\#revisions} \\ \hline
All editors found in discussions & 31,196 & 1,397,565 & -           \\ \hline
Registered                & 21,995 & 1,012,483 & 750,126,203 \\ \hline
Unregistered            & 9,130  & 62,976    & -           \\ \hline
bots                             & 71     & 322,106   & -           \\ \hline
\end{tabular}
\end{table}

\subsection{Data preprocessing}
\begin{figure}[h]
\centering
\includegraphics[width=\linewidth]{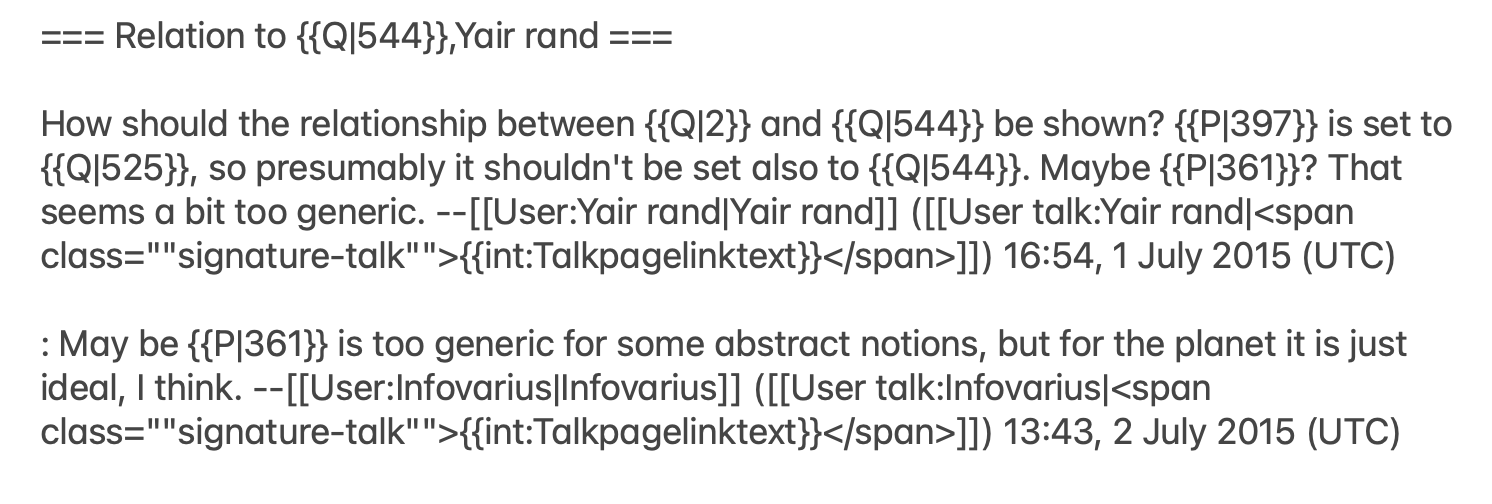}
\caption{An example of a raw Wikidata thread from the item ``Earth''.}
\label{fig:back_talk}
\end{figure}

During the data preparation, we processed raw text data. Figure \ref{fig:back_talk} shows an example of raw text from a discussion page and Figure \ref{fig:front_talk} the respective front page for the same discussion. In the figure, we can see that raw text follows a template for every page using specific punctuation to identify the subjects of the threads(i.e., {\fontfamily{qcr}\selectfont === subject ===)}, and the username (i.e., {\fontfamily{qcr}\selectfont [[User: username|username]]}) and timestamp (i.e., {\fontfamily{qcr}\selectfont hour: minutes date month year (UTC)}) at the end of each post. 
Similar to previous studies in Wikipedia \cite{viegas2007talk,massa2011social}, we phased multiple challenges in the process of the raw text discussions. We processed the text taking place on a discussion page, starting with separating threads and posts on the page, and thereafter extracting the information on the contributors who wrote the posts, and when. Due to the large size of the Wikidata archived dumps, $1.6T$, the minor changes in templates that happened in the project over the years, the differences in terms of discussion channel (e.g., some channels include multiple table templates, others change their header/subject type over time), and human error, we expected a small percentage of error in our dataset when separating threads and posts, and misspellings in the editors' usernames. 

We encountered three types of difficulty related to usernames: some posts missed the expected username signature at the end, probably due to editors who were inexperienced with Wikidata practices (this is similar to Wikipedia for $8.3\%$ of posts \cite{massa2011social}); the multilingual nature of the projects made it challenging to follow the structure of the template in all languages and track the username in raw text; and the freedom provided by Wikidata to create usernames with colours, highlights, and emojis made it challenging to identify the usernames due to the ASCII encoding format related to this features that we found in the raw text.

\section{Methodology}
\label{sec:methodology}

We created a set of features for Wikidata editors similar to studies related to the behaviour of contributors in peer-production communities.
Previous research on Wikidata examined the duration of editors' participation \cite{sarasua2019evolution}. We introduced the feature \textit{account age} to analyse the various membership lengths in months that editors have within this community, and the \textit{activity status} to distinguish the active editors from those dropped out. Furthermore, in Wikidata and Wikipedia, previous studies analysed the volume of edits for the editors \cite{brandes2009network,kittur2007he,sarasua2019evolution}. 
Similarly, we included the features \textit{number of edits}, and \textit{number of posts}. In ontology engineering studies, such as Schema.org \cite{kanza2018does}, the authors analysed the characteristics of editors. In this vein, we introduced the \textit{talking type}, which determines whether an editor mostly starts a discussion or responds to others, and the \textit{access level}, which specifies the role of editors in the community.
Finally, in the case of MSN and Slashdot \cite{leskovec2008planetary,gomez2008statistical}, the authors studied network metrics for their user networks, which led us to study centrality metrics, such as \textit{degree, eigenvector, closeness, betweenness}, and \textit{clustering coefficient}.

Table \ref{tab:feature_descritpion} shows the features and their descriptions. The history of edits dataset was used to measure each editor's \textbf{account age}, which is the gap between the day they registered and their last edit, and the \textbf{number of edits}. For registered days, we used a Wikimedia API.\footnote{\url{https://www.mediawiki.org/wiki/API:Main_page}} The discussions dataset was used to count the \textbf{number of posts} and classify editors into `asker' and `responder'. 

\begin{table*}[h]
\caption{The lists of features used to investigate the behaviour of editors and their descriptions.}
\scriptsize
\centering
\label{tab:feature_descritpion}
\begin{tabular}{|p{0.1\linewidth}|p{0.85\linewidth}|}
\hline
\textbf{Feature}       & \textbf{Description} 
\\\hline
account age            & The range in months between the editor’s registration date and their last edit timestamp.                                                   \\ \hline
\# of edits            & The total number of edits excluding discussion page edits. 
\\ \hline
\# of posts            & The total number of posts in discussion pages.                                                                                                \\ \hline
talking type           & Label related to the way editors talk: `asker' for editors mostly starting threads; `responder' for editors mostly responding to threads. \\ \hline

access level &
  The role/roles of editors based on Wikidata hierarchy (Basic level users, Rollbackers, Translation administrators, Administrators, IP   block exceptions, Property creators, Confirmed users, Interface administrators, CheckUsers, Wikidata staff, Bureaucrats, Suppressors, Flooders) \\ \hline
degree                 & The number of edges for a node.                                                                                                               \\ \hline
eigenvector &
  Defines the centrality of a node as proportional to its neighbours’   importance. For example, for two nodes in a network with the same degree, the one connected to more nodes with a high eigenvector score should have a higher eigenvector score. \\ \hline
closeness              & Counts the average of shortest paths. Nodes with a high closeness score have the shortest distances to all other nodes.                  \\ \hline
betweenness            & Based on the shortest path between nodes. Counts the number of times a node acts like a bridge.                                           \\ \hline
clustering coefficient & Quantifies how close its neighbours are to be a clique (complete graph).                                                              \\ \hline
activity status &
  Label related editors’ current activity: active for editors   with range between their last edit and today (1/3/2023), less than 23 months;   inactive for editors with the same range more than 23 months. \\ \hline
\end{tabular}%
\end{table*}

For the feature \textbf{access level}, we visited the Wikidata list of users page,\footnote{\url{https://www.wikidata.org/wiki/Special:ListUsers}} and searched each access level group individually. We then assigned labels to our list of editors using the list of usernames for each access level.

Our feature list includes five centrality metrics, which are \textbf{degree, eigenvector, closeness, betweenness} and \textbf{cluster coefficient}. These metrics were measured based on the registered editors' network, described in detail below, Section \ref{meth:RQ1}. 

\begin{figure}[b!]
\centering
\includegraphics[width=0.7\linewidth]{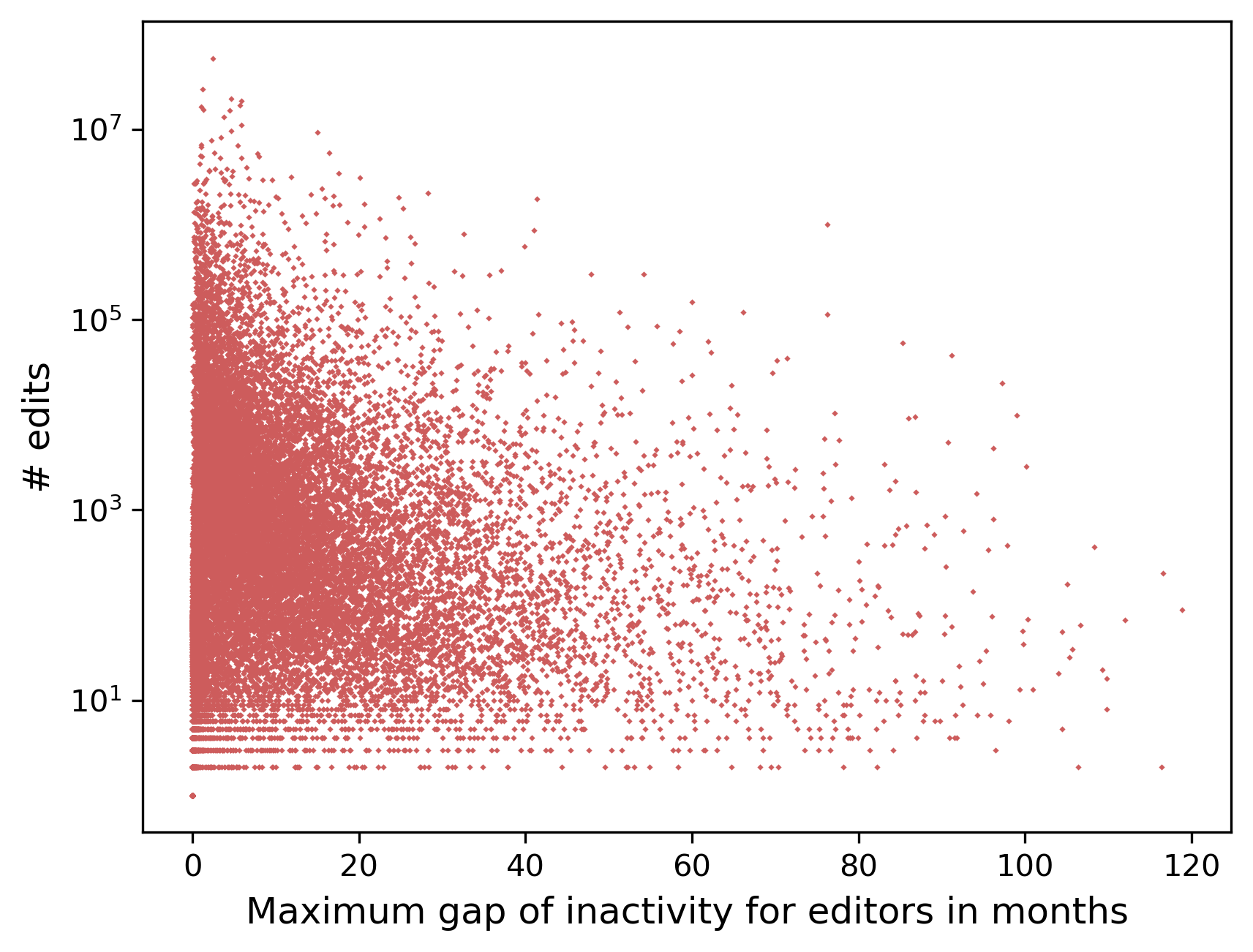}
\caption{The maximum gap of inactivity in months and the number of edits for registered editors discussing in Wikidata.}
\label{fig:gaps}
\end{figure}

Finally, we built a notion about who was considered an active or inactive editor for the label \textbf{activity status}. Wikidata classifies editors as active and inactive based on their monthly editing activities without considering editors permanently leaving or dropping out. In contrast,  
Sarasua et al. \cite{sarasua2019evolution} highlighted that editors presented varying periods of absence with no edits, leading to a comeback or drop out. Therefore, the authors used the percentiles of editors for gaps of inactivity and defined that after approximately $10$ months of no contributions, editors were considered permanently inactive. Similar to Sarasua et al. \cite{sarasua2019evolution}, we used the percentile of editors for the maximum gaps of inactivity of editors. 
Based on Figure \ref{fig:gaps}, we can see that the majority of editors had a maximum gap of inactivity of less than $20$ months. We chose the $85\%$ percentile, defining a threshold of $23$ months. According this definition, we found that $67\%$ of editors were active and $33\%$ inactive.

During the ten years of Wikidata, we found several editors who were absent for five to nine years but returned to contributing (see Figure \ref{fig:gaps}). The maximum gap of inactivity was $118$ months. These findings differ from the first four years of Wikidata, where Sarasua et al. \cite{sarasua2019evolution} found the maximum gap of inactivity being $16$ months and $55\%$ of editors inactive. There are numerous reasons, from personal issues to global events, why editors present these long absences. Further analysis could reveal insights into engagement patterns.  
It is interesting to note that editors with long periods of inactivity present a variety in the number of edits. Figure \ref{fig:gaps} shows that editors with almost seven years of absence can have from $10$ to $100K$ edits.

\subsection{RQ1 - What are the characteristics of editors’ collaborations?}
\label{meth:RQ1}

We conducted network analysis and descriptive statistics. Network analysis is used across various fields due to its ability to reveal complex relationships and patterns within data. Its ability to reveal hidden patterns, optimize network performance, and identify key nodes makes it an invaluable tool across a wide range of disciplines. We built and analysed three editor networks: (i) with all editors found in the discussions dataset (NF); (ii) with human editors, registered and unregistered (NH); and (iii) with only registered editors (NR). 
In each network, we considered nodes as editors (humans or bots) who had at least one post on a Wikidata discussion page. We added edges between editors participating in the same thread. 

Wikidata bots, like Wikipedia bots, participated in discussions either through bot controllers or by delivering automatic messages \cite{massa2011social}. We elaborate in Section \ref{sec:compare_networks}. 

We employed centrality metrics (see Table \ref{tab:feature_descritpion}) \cite{bhattacharya2020impact} to investigate and compare the three editor networks. Centrality measures help to \textbf{explore the characteristics of the Wikidata network of editors}. They identify the most important or influential nodes in the network and explain the connection of nodes. This can be critical for targeting interventions or understanding power dynamics.

We then further explored the behaviour of editors in the NR network. Specifically, we investigated whether NR forms a small world network by analysing the connected components in the network (which is a connected sub-network that is not part of any other larger network connected component) and measuring the average clustering coefficient and average shortest path. This analysis helps to \textbf{explore the communication characteristics of the registered editors}. It examines the robustness of the network and the connectivity of groups. 
A small world network forms a giant component and presents a high average clustering coefficient and low average shortest path \cite{watts1998collective}. The clustering coefficient (as described in Table \ref{tab:feature_descritpion}) measures the degree to which nodes in a network cluster together. It quantifies the likelihood that two neighbours of a node are also connected to each other. The shortest path refers to the minimum number of steps or edges required to traverse from one node to another node in a network.
To test our small world hypothesis, we compared these metrics with a similar small world network generated with the Watts-Strogatz method \cite{watts1998collective}, as well as with a similar random network. 

We then explored network characteristics like the distribution of clustering coefficient, shortest paths, and assortativity \cite{newman2003social}. This helps to understand the structural tendencies of the network, indicating whether it is more segmented into active subgroups or more integratively connected.


Finally, we analysed the characteristics of registered editors using the features listed in Tables \ref{tab:feature_descritpion}. This analysis can provide insights into \textbf{the structural properties of the network of editors} to explore the behaviour of editors and engagement patterns.

\subsection{RQ2 - What factors affect whether a discussion will receive a response?}

We developed a machine learning model to study the factors that affect discussions in Wikidata. The aim was to predict the likelihood of a response to a post based on various features. We combined features of: characteristics of the editors who wrote the post; graph embedding which are a type of representation learning where a graph is converted into a low dimensional space in which the graph information is preserved \cite{cai2018comprehensive}; and text embedding, which are numerical representations of text that capture the semantic meaning of words, phrases, sentences, or entire documents \cite{patil2023comparative}. The list of features used for the ML model is presented in Table \ref{tab:ML_features}.
We conducted an ablation study, which means systematically removing or masking individual features (or groups of features) and observing the change in model performance. Then, we compared the performance metrics (e.g., accuracy and precision) of the model before and after removing each feature. This analysis helped us to understand which factors, either individually or in combination, affect the prediction of the posts.

\begin{table}[t]
\caption{The list of features used in the ML model with descriptions.}
\label{tab:ML_features}
\begin{tabular}{|l|p{0.8\linewidth}|}
\hline
\textbf{Feature} & \textbf{Description}                                                    \\ \hline
EFage   & Account age: The range in months between the editor’s registration date and their last edit timestamp.                                         \\ \hline
EFedits          & \# of edits: The total number of edits excluding discussion page edits. \\ \hline
EFposts          & \# of posts: The total number of posts in discussion pages.             \\ \hline
EFaccess level   & Access level: The role/roles of editors based on Wikidata hierarchy.    \\ \hline
EF               & All the above features.                                                 \\ \hline
Gpost            & Graph embeddings for posts.                                             \\ \hline
Geditors         & Graph embeddings for editors.                                           \\ \hline
TfirstPost       & Text embeddings for the first posts of each thread.                     \\ \hline
Tpost            & Text embeddings for individual posts                                   \\ \hline
Tthread & Text embeddings for each post combining the previous posts in the thread. This is the average of text embeddings for the first to the ith post \\ \hline
Tavr    & Text embeddings for the editors. This is the average of the text embeddings for all their posts.                                               \\ \hline
\end{tabular}
\end{table}

For the \textbf{features of editors} (EF), we used the set of features (Table \ref{tab:feature_descritpion}) proposed in Section \ref{meth:RQ1} for the investigation of the characteristics of editor's collaboration.
We used account age (EFage), number of edits (EFedits), number of posts (EFposts), and access level (EFaccess level).
 
\begin{figure}[h]
\centering
\includegraphics[width=0.5\linewidth]{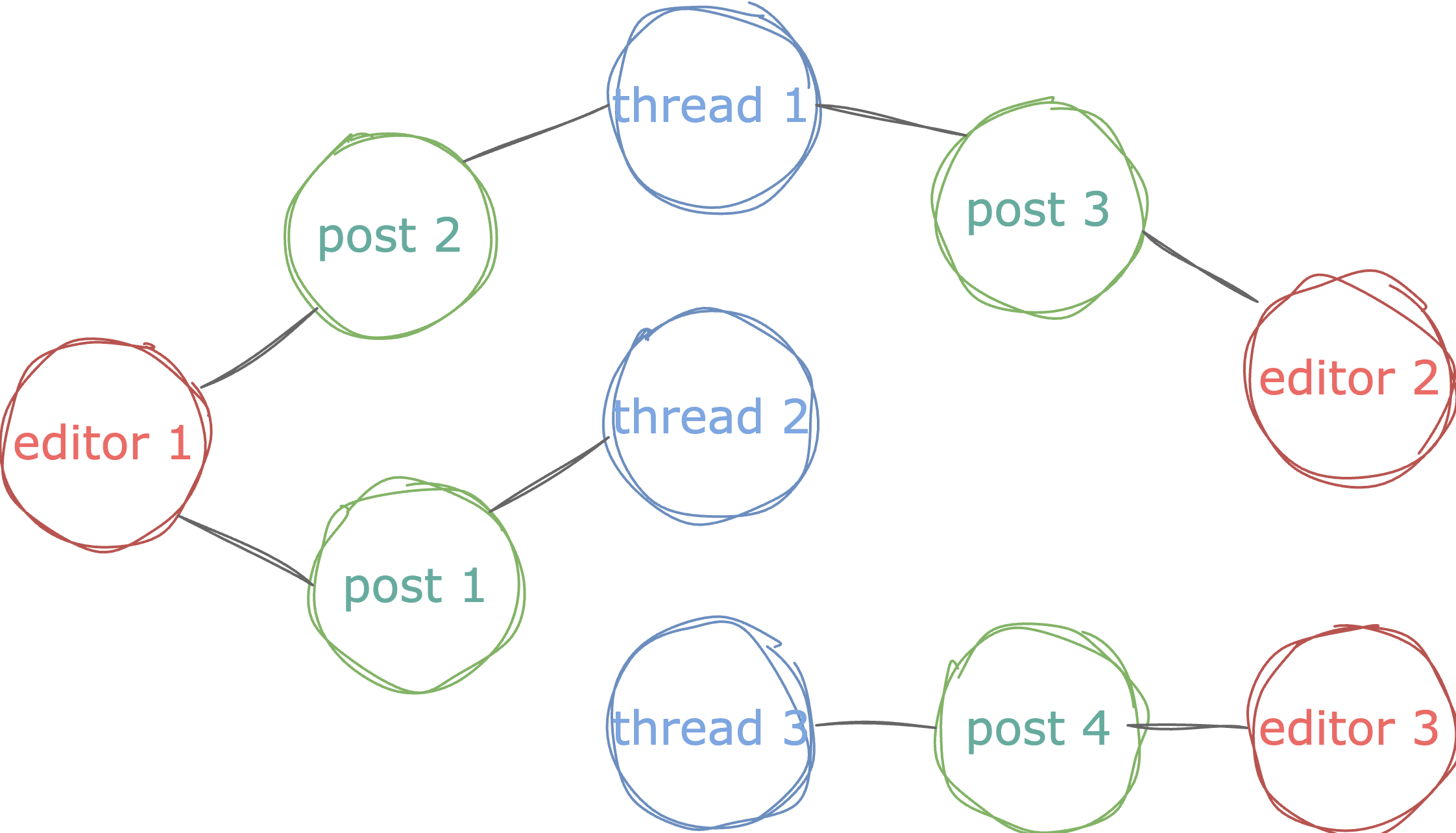}
\caption{Representation of the network to estimate graph embeddings.}
\label{fig:network}
\end{figure}

For \textbf{graph embeddings} (G), we created a discussion network including three types of nodes: editors, posts, and thread titles. Editors were connected to their posts, while posts were connected to the threads they belonged (Figure \ref{fig:network}). 
The graph included $75,055$ nodes ($6,596$ editors, $44,139$ posts, $24,320$ threads) and $88,092$ edges.
To create the graph embedding vectors for this network, we experimented with several types of state-of-the-art models. 
We experimented with translational distance models like TransR \cite{lin2015learning}, RotatE \cite{sun2018rotate}, which employs rotations in the complex vector space, and the CompGCN \cite{vashishth2019composition}, which combines the principles of graph convolutional networks with compositional operators.
Due to the network structure and the low number of relations, none of the tested models presented good performance except for CompGCN. For CompGCN we applied embedding dimensions equal to $100$ to form a graph embedding vector for each entity in this network (i.e., editors, posts, and threads). CompGCN showed the best performance with Hits@1 $0.37$, Hits@3: $0.56$, Hits@5: $0.64$, Hits@10: $0.75$ and Mean Reciprocal Rank $0.5$. 

The full discussion network was composed of more than $2.5M$ edges, which required high computational power to estimate the graph embedding vectors. Therefore, we used only discussions found in item talk pages as a sample and not the whole discussion dataset as described in Section \ref{sec:data}. We chose this sample because item talk pages presented the best balance between posts receiving a response and those not ($44\%$ of posts had a response, $56\%$ did not). Our previous study on Wikidata discussions also showed that in item talk pages, only $50\%$ of posts starting a new thread receive community attention to get an answer \cite{koutsiana2023analysis}. In addition, as shown in Koutsiana et al. \cite{koutsiana2023analysis}, item talk pages include a wide range of topics and issues, such as KE topics, fact accuracy, conflicts, and policies. These topics can reflect better than other discussion channels, such as property talk pages and project chat, the conversations across Wikidata \cite{koutsiana2023analysis}.

To answer $RQ2$ we used the graph embedding vectors generated for the posts (Gpost).

For \textbf{text representations} (T) we used the $text-embedding-ada-002$, a $GPT3.5$ large language model from OpenAI. We used this model because it was among the best performing models in the literature, and it was capable of processing long documents like post discussions effectively \cite{patil2023comparative}. 
We generated two types of text embeddings: one for each post (Tpost) and one for the entire thread from the first to the \textit{i}th post (Tthread). Prior to generating these embeddings, we preprocessed the raw data by removing signatures and timestamps at the end of the post, replacing the item and property alphanumeric codes ({\fontfamily{qcr}\selectfont Q\#} and {\fontfamily{qcr}\selectfont P\#}) with {\fontfamily{qcr}\selectfont [ITEM]} and {\fontfamily{qcr}\selectfont [PROP]}, respectively, and any code elements with {\fontfamily{qcr}\selectfont [URL]} or {\fontfamily{qcr}\selectfont [CODE]}. Finally, we removed any remaining unnecessary punctuation.

\begin{figure}[h]
\centering
\includegraphics[width=\linewidth]{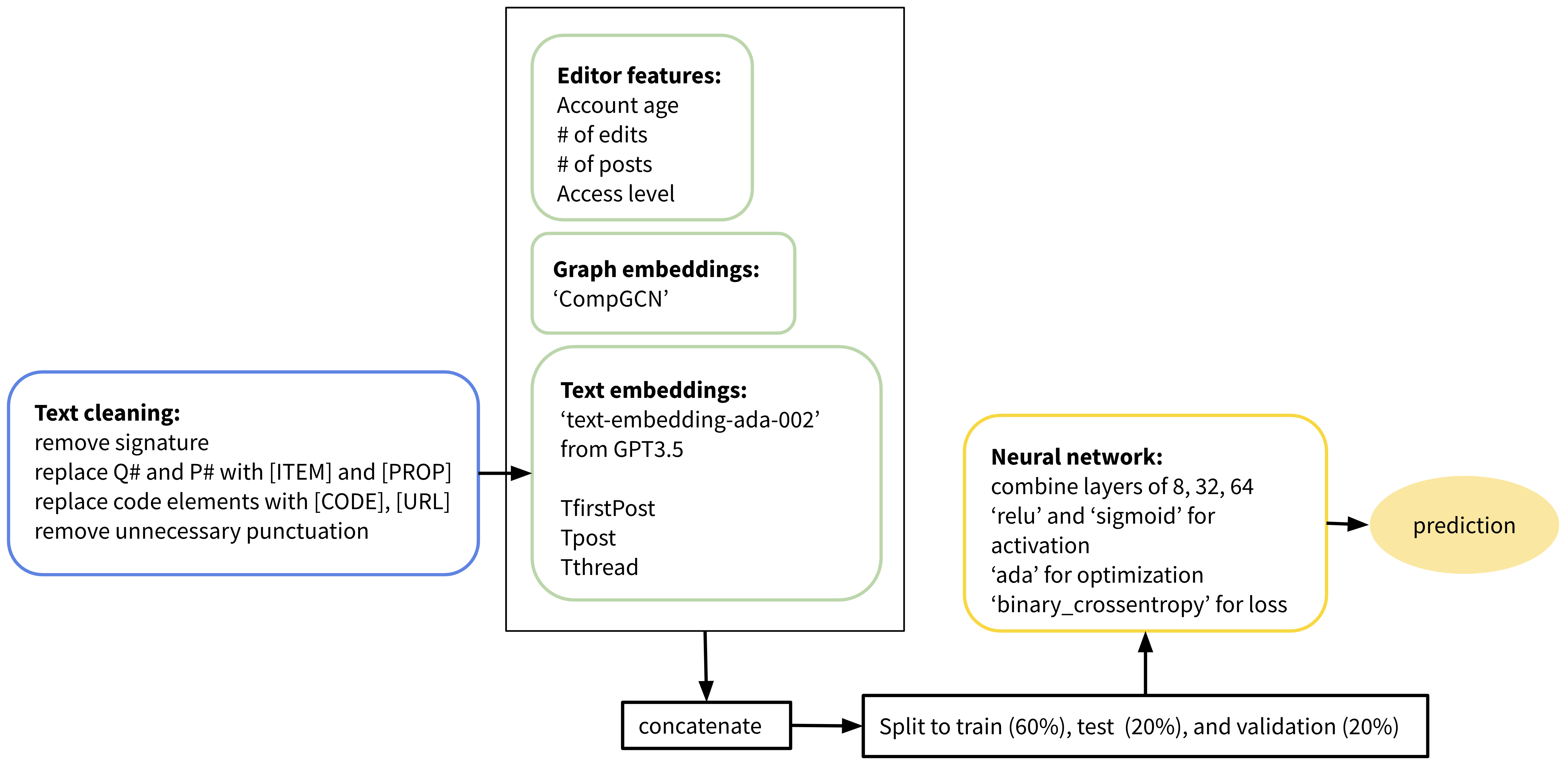}
\caption{A diagram with the steps for the ML model.}
\label{fig:ML}
\end{figure}

To develop the \textbf{machine learning model}, we used a neural network and concatenated the above features.
Figure \ref{fig:ML} illustrates the steps of the ML model.
We used a Multi-Layer Perceptron (MLP) because it is a versatile, efficient, and straightforward choice for a wide range of machine learning tasks, particularly when dealing with structured data and simpler relationships \cite{khalil2009performance}. 
To train the model, we randomly split our dataset to train, test, and validate sets of $60\%$, $20\%$, and $20\%$, respectively. 
The MLP had $1\-3$ layers (test combinations of $8$, $32$, and $64$ neurons) with ``relu'' as activation function. The output layer consisted of a single neuron with ``sigmoid'' as an activation function. For training the network, we used ``adam'' as the optimizer with a learning rate set at $0.001$ and the ``binary\_crossentopy'' loss. Furthermore, we used early stopping to stop the training if the validation loss did not improve in $100$ epochs. We evaluated results using precision, recall, accuracy, and F1.

\subsection{RQ3 - Do discussions affect the editors' engagement?}

To address the $RQ3$, we employed the same approach as in $RQ2$. We used the same ML model (Figure \ref{fig:ML}). This time, we delved into the factors that affected the engagement of editors, which we defined using the feature activity status as described above (see Table \ref{tab:feature_descritpion}). This feature indicated whether an editor was \textit{active} or \textit{inactive}.

As in $RQ2$, we leveraged various features for editors Table \ref{tab:ML_features}), including their graph embedding vectors (Geditors), and text embedding vectors for their posts. For the latter, we estimated the average of text embeddings for all posts of each editor (Tavr).We then concatenated all these features to the ML model (Figure \ref{fig:ML}). We evaluated results using precision, recall, accuracy, and F1.

To validate our findings and determine the statistical significance of the features, we utilized a range of statistical tests depending on the particular case.

\section{Results}
\label{sec:results}
\subsection{RQ1 - What are the characteristics of editors’ collaborations?}

\subsubsection{Comparison of the three editor networks}
\label{sec:compare_networks}

\begin{figure}[h]
\centering
\includegraphics[width=0.7\linewidth]{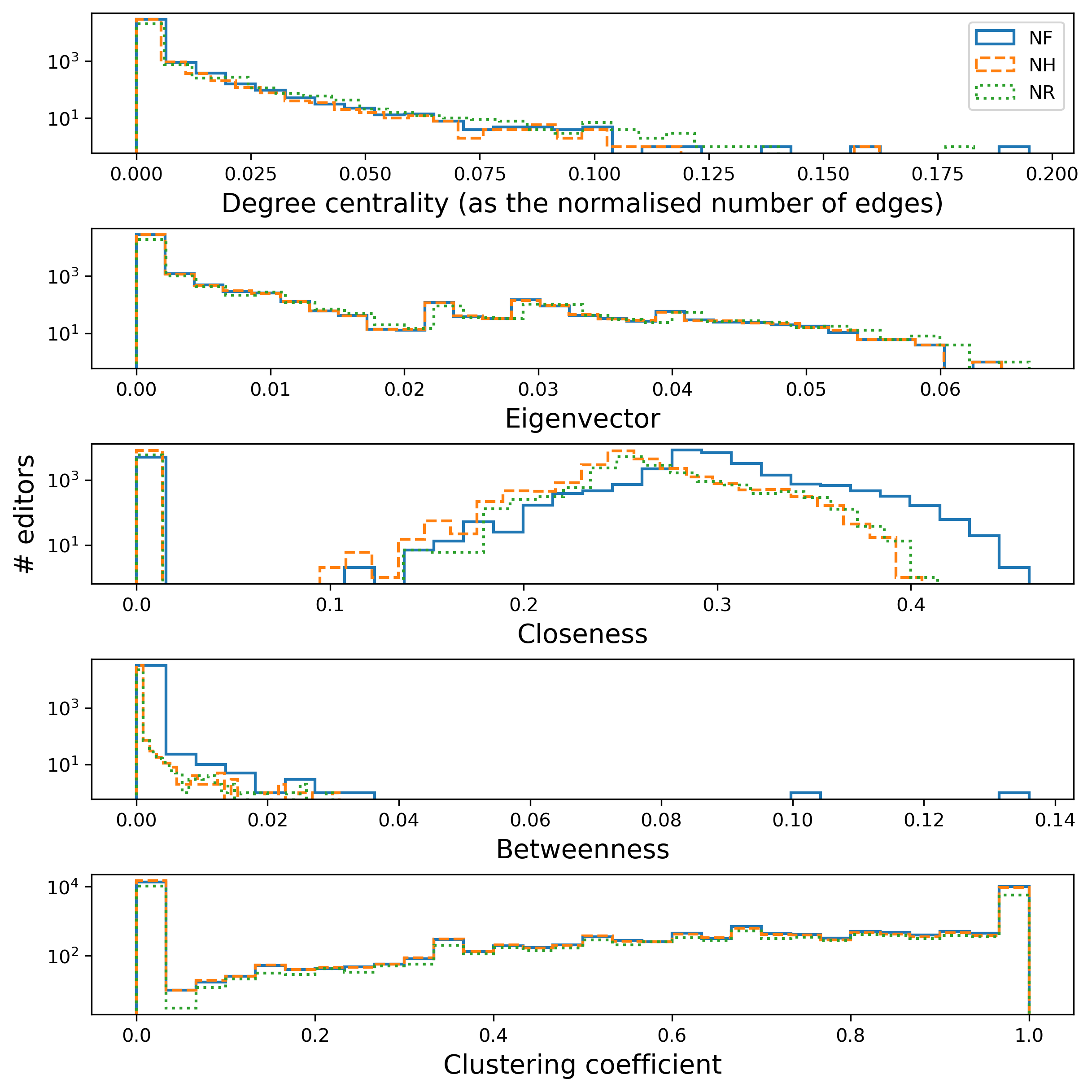}
\caption{Centrality metrics for the three networks}
\label{fig:centrality}
\end{figure}

Figure \ref{fig:centrality} presents centrality metrics for the NF, NH, and NR networks.
For all networks, \textbf{degree centrality}, which measures the number of edges for a node, presented a power law distribution with a high number of editors with no edges in all networks: $14\%$ NF, $24\%$ NH, $26\%$ NR. This indicated that a quarter of human editors did not receive responses when starting a discussion. Furthermore, $47\%$ of them were inactive, meaning that half of the editors who did not receive a response drop out, while the rest continued contributing without discussing. In contrast, $68\%$ of editors who received a response are still active. 
To further investigate this relation, we analysed whether the activity status of an editor was related to their posts received responses using the Chi-square test for categorical variables \cite{ugoni1995chi}. We found a p-value less than 0.05, which showed that these two features were related.

For \textbf{eigenvector} and \textbf{clustering coefficient} metrics, the three networks presented similar distributions (Figure \ref{fig:centrality}). The eigenvector shows the centrallity of a node as proportional to its neighbours' importance, and the clustering coefficient quanitfies how close its neighbours are to a complete graph.
We did not observe any nodes with superior or inferior connections or clustering for registered editors, unregistered editors, and bots. This means that the removal of bots or unregistered editors did not significantly affect the network’s overall structure and influence distribution. The network might have a high degree of connectivity and robustness, where the influence and connections are well-distributed among many nodes. This makes the network resilient to the removal of nodes. We explore this further below in Section \ref{sec:reg_editors}.

In terms of \textbf{closeness}, which counts the average of shortest paths, we observed that NF nodes had shorter distances (high closeness), while NH and NR showed similar distribution.
At the same time, \textbf{betweenness}, which counts the number of times a node acts like a bridge, presented lower values for NF.
The higher closeness values and lower betweenness values for NF suggested that bots were the most connected editors in the network and helped in bridging distances between nodes, thereby making paths for all nodes shorter. However, the eigenvector and clustering coefficient showed that even when removing bots, the network stays resilient.

After the analysis, we looked into the nodes with the highest centrality metrics for the three networks. 
In the NF network, a bot had the highest degree, closeness, and betweenness values.  
``DeltaBot'' was the most active editor in discussions. It responded to requests for deletion or merging of items with prefixed answers such as ``\{\{on hold\}\} This item is linked from 3 others'' after checking their status. However, as described in Section \ref{sec:background}, the role of bots is to perform repeated tasks, and the purpose of this communication is to respond to requests with a single post. Bots do not interact with editors or participate in conversations.    
For NH and NR networks, editors with administrative access levels, such as property creators or administrators, received the highest metrics. 
The analyses indicated the importance of bots in communication related to maintenance jobs, and the dedication of the editors with higher level access to communication.

\subsubsection{Investigation of registered editor network}
\label{sec:reg_editors}

We checked whether the communication interactions of Wikidata editors formed a small world network.
We first examined the connected components for the NR network. We discovered that there were $5,722$ connected components, with the largest component containing $73\%$ of the nodes, which amounts to $16,169$ nodes and $470,126$ edges. The second-largest component included only $0.04\%$ of the nodes. The NR network had a giant component, and this was the first indication of a small world network. 

Next, to investigate if Wikidata was a small world network, we measured and analysed the average cluster coefficient and the average shortest path of NR. The average clustering coefficient was $0.6$ and the average shortest path was $3$. To compare the characteristics of the Wikidata network, we generated a small world network and a random network with a similar number of nodes and edges. In comparison, a similar small world network had similar characteristics with an average clustering coefficient of $0.1$ and an average shortest path of $3$. In contrast, a random network with similar characteristics had a much lower average clustering coefficient of only $0.004$ and an average shortest path of $3$. These comparisons, combined with the giant component, confirmed that NR formed a small world network. Information or resources could be disseminated quickly and efficiently across the network due to the short path lengths. Even though nodes were part of sub-groups, they could still reach other parts of the network with relatively few steps. Small-world networks are generally robust to random node removal. In the Wikidata case, this made it resilient to drop outs.

We then measured degree assortativity to investigate the tendency of editors to connect with editors with similar degree metric. NR network had assortativity at $-0.2$. This suggests a disassortative network, meaning that editors with high degree tended to connect with editors with low degree. This suggests an inclusive community with the flow of information or resources from well-connected nodes to the periphery.

\begin{figure}[h]
\centering
\begin{minipage}[]{0.49\textwidth}
\centering
    \includegraphics[width=\textwidth]{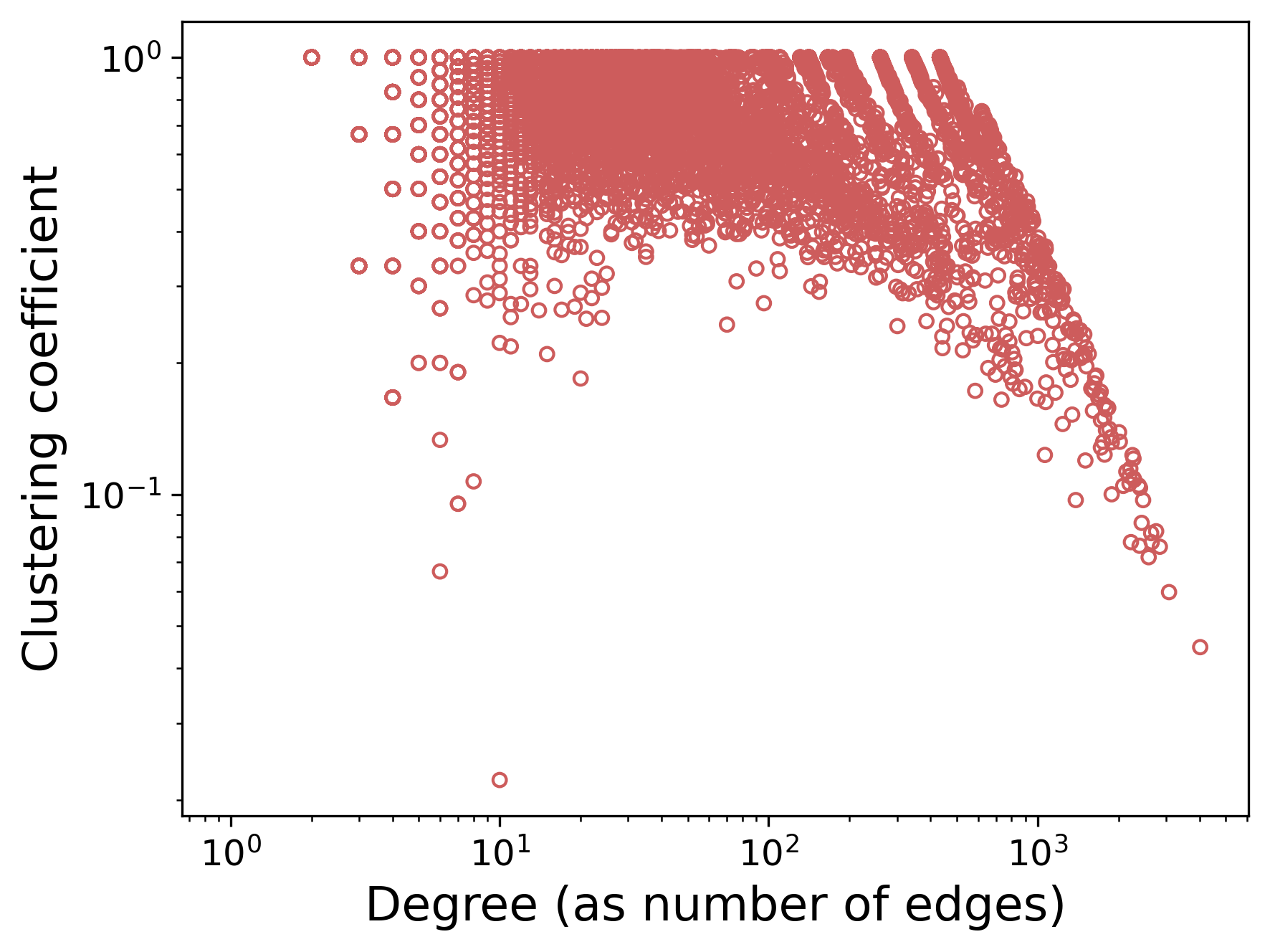}
    \caption{Clustering coefficient}
    \label{fig:cluster}
  \end{minipage}
  \begin{minipage}[]{0.49\textwidth}
  \centering
    \includegraphics[width=\textwidth]{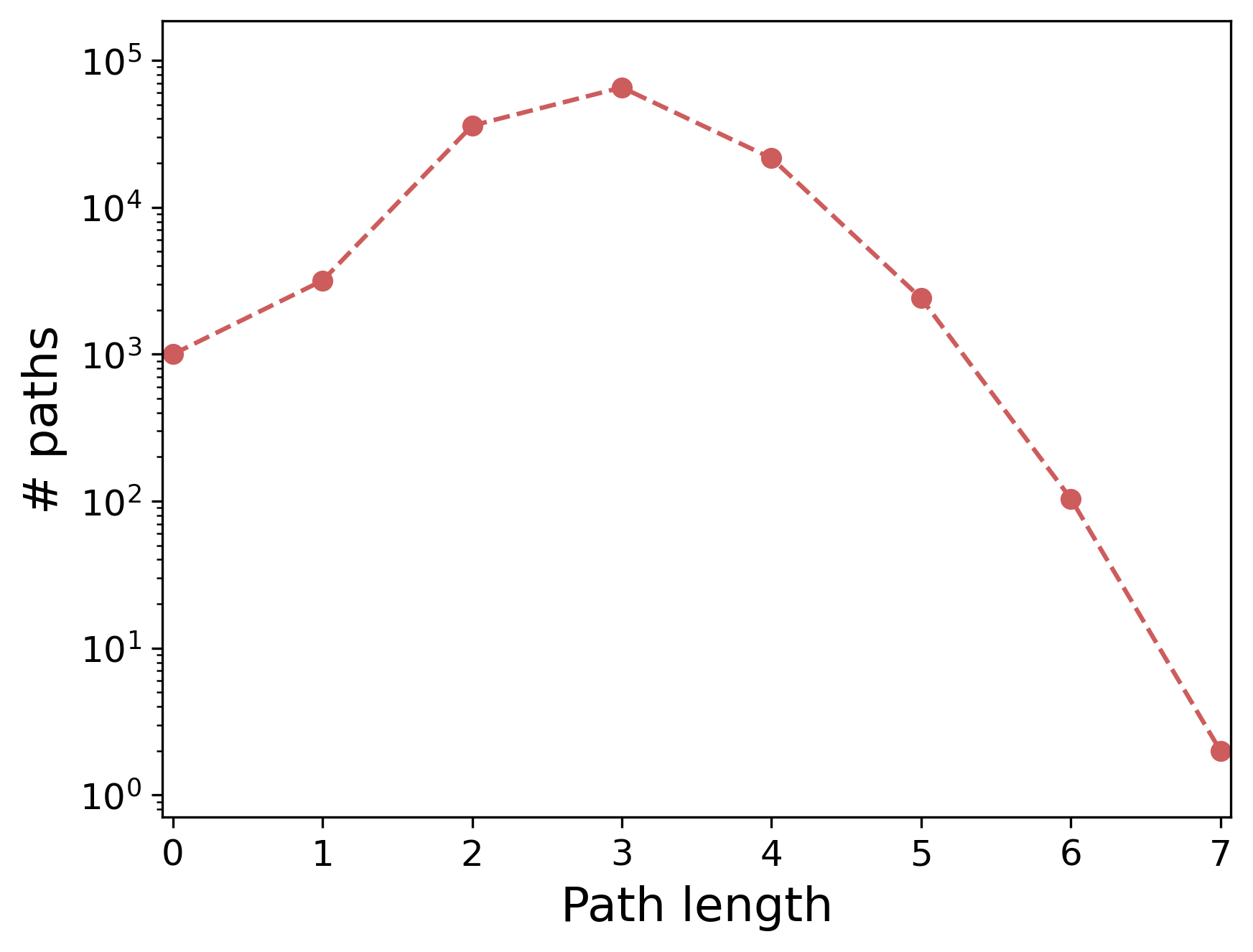}
    \caption{Shortest paths}
    \label{fig:path}
  \end{minipage}
\end{figure}

Figure \ref{fig:cluster} presents the clustering coefficient versus the degree of a node in the NR network. The figure showed that a high number of nodes had both high clustering coefficient and degree. Such nodes were likely to be influential or central within their respective clusters or communities and lead to robust networks with multiple paths for information or interactions to flow. As a result, the network was resilient to the removal of individual nodes (i.e., editor drop out). As mentioned earlier, this could be the reason why NF, NH and NR presented similar clustering coefficient distributions in Section \ref{sec:compare_networks}. Despite bots being the most connected editors in NF, their removal in NH and NR did not influence node connections.

We also measured the distribution of the shortest paths in the NR network. Similar to the MSN study \cite{leskovec2008planetary}, we randomly sampled $1000$ nodes and calculated their shortest paths to all other nodes (Figure \ref{fig:path}). We observed that the longest shortest path was seven, which indicates that long paths did not exist in the NR network. By examining the percentile of shortest paths, we discovered that one can reach $90\%$ of nodes in just four steps. This indicates that the network was highly interconnected with short average path lengths. This reflects efficient communication, robustness, and the small-world nature of the network, where local clustering and global reach coexist.

\subsubsection{Investigation of registered editors characteristics}

\begin{figure}[h]
\centering
\includegraphics[width=0.7\linewidth]{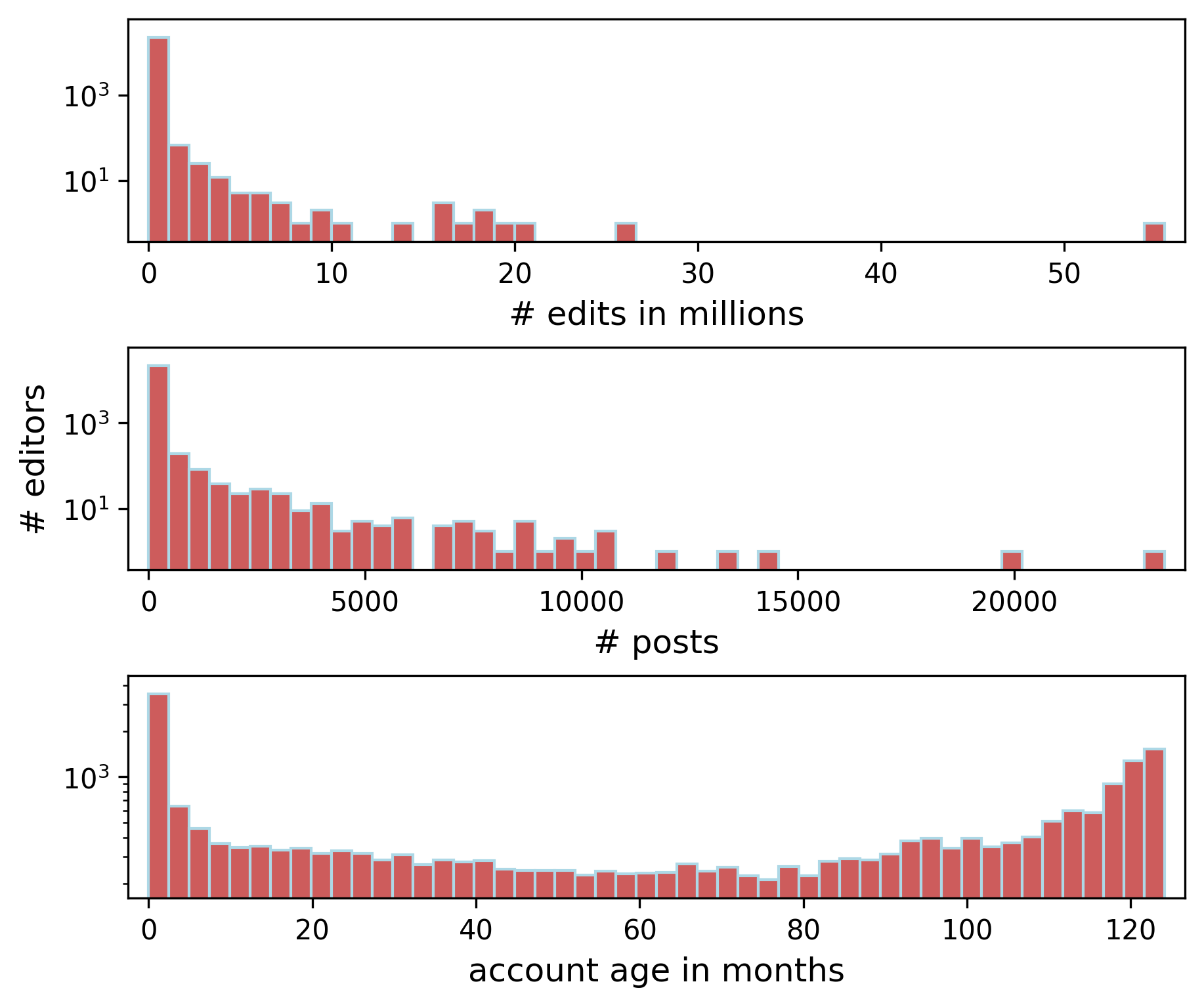}
\caption{Frequency of editors versus their features, number of edits, number of posts and account age.}
\label{fig:histograms}
\end{figure}

Figure \ref{fig:histograms} illustrates the distribution of registered editors based on the number of edits, posts and account age.
Our analysis revealed that both the number of edits and posts followed a power law distribution. 
Approximately $80\%$ of editors had fewer than $5$K edits and ten posts, suggesting low engagement with the community. Only a small percentage of editors ($20\%$) were actively participating in discussions, suggesting that they were potentially the ones making the key decisions within the Wikidata community.

\begin{figure}[h]
\centering
\includegraphics[width=0.8\linewidth]{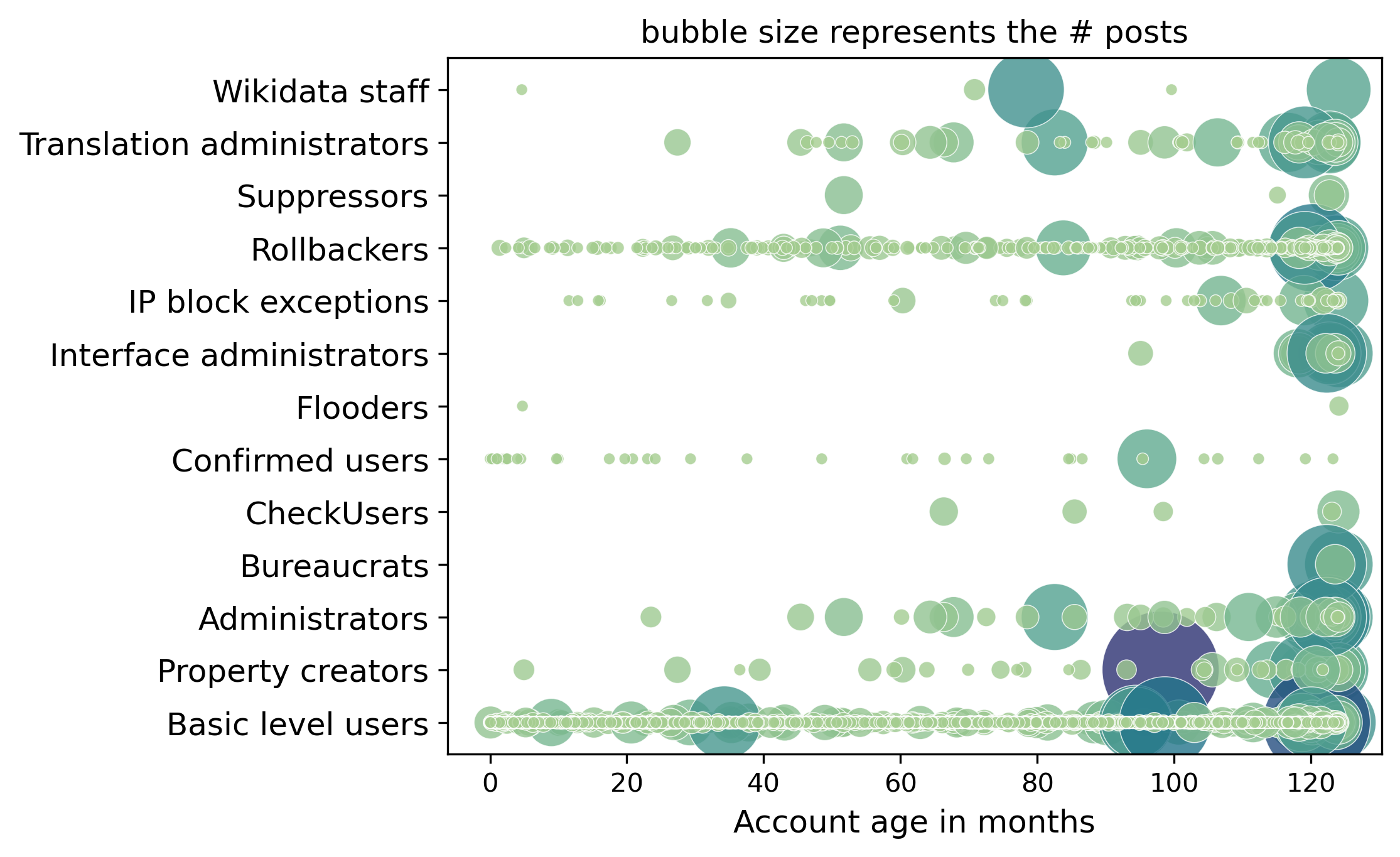}
\caption{Number of posts and the account age for the different access levels.}
\label{fig:bubble}
\end{figure}

\begin{figure}[h!]
\centering
\includegraphics[width=0.8\linewidth]{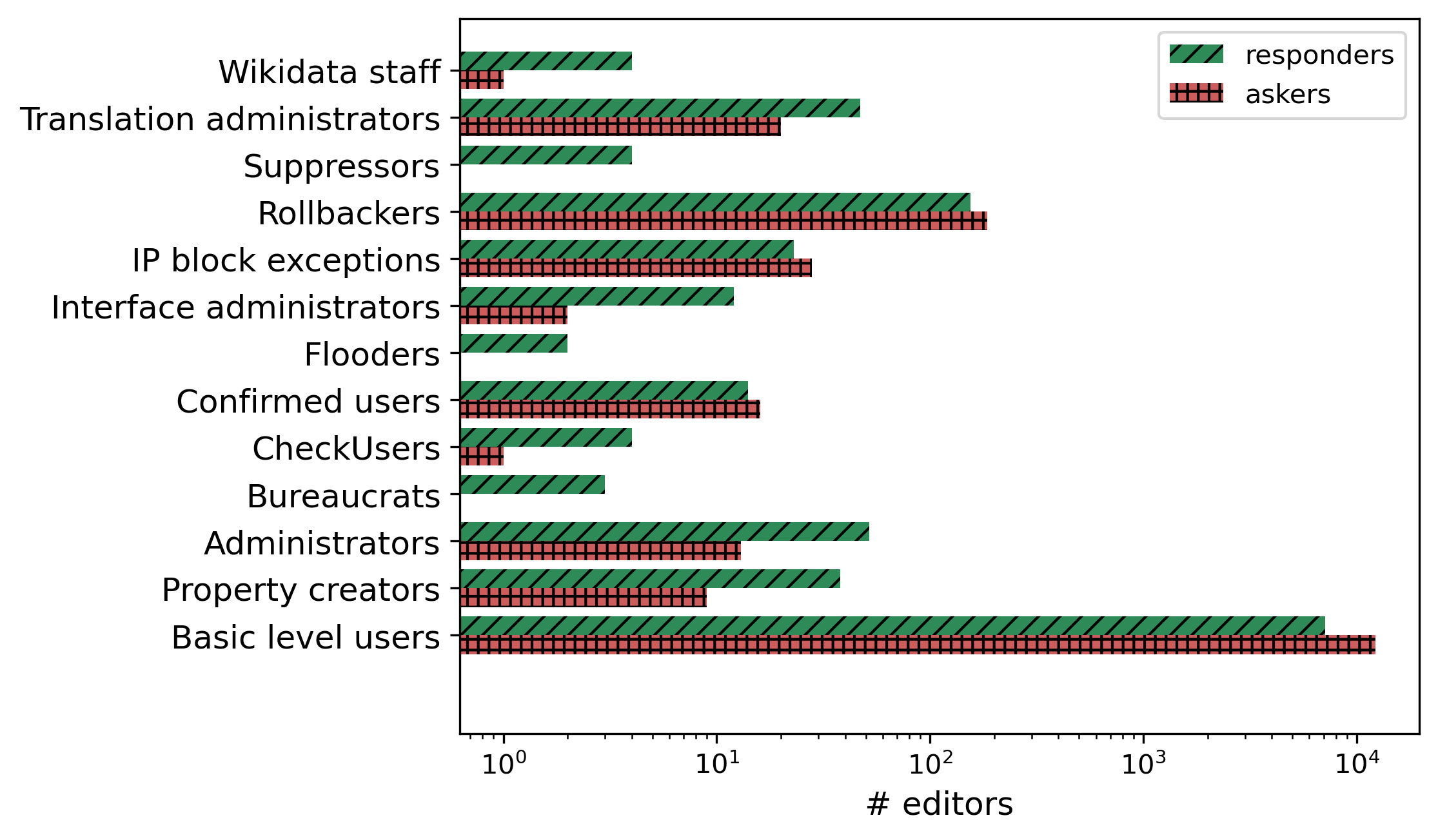}
\caption{The histogram presents the number of editors and the talking type for the different access levels.}
\label{fig:right_talking}
\end{figure}

Furthermore, we analysed the distribution of account age and found that $24\%$ of editors had stayed in the community for less than a year, and $10\%$ of editors had remained active for more than ten years. This trend was consistent with the first four years of Wikidata \cite{sarasua2019evolution}.

We further determined which editors made high contributions to discussions based on their access level and account age (Figure \ref{fig:bubble}). Editors who had been active on Wikidata for over six years were the most active in discussions.
\textit{Property creators} were the most active, followed by \textit{basic level users}. This suggests that editors with basic level editing rights made a great share of contributions in discussions, such as helping others and participating in decisions, etc.

We then explored which access levels usually started a thread,\textit{askers}, and which ones usually responded, \textit{responders}, to posts.
Figure \ref{fig:right_talking} shows that most of the editors participating in discussions were \textit{basic level users} followed by \textit{rollbackers, translator administrators}, and \textit{administrators}.
The majority of \textit{basic level users} and \textit{rollbackers} started threads, while administrative roles such as \textit{administrators} and \textit{bureaucrats}, and KG experts such as \textit{property creators}, were responders. This was not surprising given that \textit{basic level users} often seek advice, request access, and tool support from the community, particularly from members with higher access levels.

\subsubsection{Summary} 
According to our findings, around a quarter of registered editors were disconnected from the network, which means they did not receive responses to their posts. Further investigation showed that half of them are inactive today. In contrast, bots were the most connected editors. 

Our study also revealed that the Wikidata editor network forms a small world network, with a random node can reach $90\%$ of the nodes in just four steps. 

Furthermore, we found that only $20\%$ of editors were actively involved in discussions and decision-making, and $10\%$ of editors were members for more than six years. It is interesting to note that the majority of editors who participated in discussions were basic level users, and some of them had the highest number of posts.

\subsection{RQ2 - What factors affect whether a discussion will receive a response?}

The study aimed to investigate the factors that influence whether a post on item talk pages would receive a response.
The ablation study included text embedding for the post content, graph embedding for the post topology (Gpost), and features for the editor who wrote the post (Table \ref{tab:ML_features}). 
  
We conducted an experiment using different text embeddings to determine their effectiveness. We tested whether using text embeddings for only the first posts of threads (TfisrtPost) performed better than other posts within the threads (Tpost), or the entire thread from the first to the \textit{i}th post (Tthread). If TfirstPost and Tpost showed different performances, it could indicate that the position of the post influences the prediction. Additionally, if Tthread and Tpost showed different performances, it could indicate that previous posts in the thread can affect the prediction.

The results of the experiment are presented in Table \ref{tab:results_RQ2}. The results showed a balance between precision and recall scores for all features, suggesting that the ML model had balance in correctly identifying positive instances and avoiding false positives. This means that the model was not heavily biased towards one type of error. The model was likely to generalise well to new data, suggesting that it was not overfitting to the training data. This was further supported by the F1 scores, in which a score similar to precision and recall indicates a balanced and high-performing model.

Table \ref{tab:results_RQ2} demonstrates that, among individual features, graph embedding performed best, with $0.72\%$ accuracy, in predicting whether a post would receive a response. This result was expected since a post that was part of a highly connected thread had a higher chance of receiving a response.  
In addition, we found high performance, with approximately $0.7\%$ accuracy, for the text embeddings with minimal variations between the TfirstPost, Tpost, and Tthread. This suggested that the position of the posts and the previous posts did not have a significant influence, but the content of the text played a significant role in eliciting a response. 
The results mean that based on the content of the post, we could predict whether it receives a response. This was particularly important for TfirstPost results, which initiated a discussion. 

The best features for predicting whether a post would receive a response were its content and where it was posted (Gpost + TfirstPost), with $0.79\%$ accuracy. The characteristics of the editor, such as their account age and access level, did not seem to significantly impact whether the post would receive a response. 
This suggests that it was more important to focus on the content of the post and where it was placed rather than who wrote it. This shows that the community was inclusive. These results are similar to previous findings about the communication characteristics of the registered editors in $RQ1$.

\begin{table}[h]
\centering
\caption{Performance results of the ablation study for the features individually and in combination. The ML model predicted whether a post would receive a response.}
\label{tab:results_RQ2}
\begin{tabular}{|l|l|l|l|l|}
\hline
\textbf{Features}       & \textbf{Precision} & \textbf{Recall} & \textbf{Accuracy} & \textbf{F1}   \\ \hline
EFage                   & 0.54               & 0.55            & 0.55              & 0.54          \\ \hline
EFedits                 & 0.56               & 0.57            & 0.57              & 0.56          \\ \hline
EFposts                 & 0.59               & 0.58            & 0.58              & 0.57          \\ \hline
EFaccess level                & 0.55               & 0.56            & 0.56              & 0.53          \\ \hline
EF                      & 0.59               & 0.59            & 0.59              & 0.59          \\ \hline
Gpost                   & \textbf{0.72}      & \textbf{0.72}   & \textbf{0.72}     & \textbf{0.72} \\ \hline
TfirstPost              & \textbf{0.7}       & \textbf{0.7}    & \textbf{0.7}      & \textbf{0.7}  \\ \hline
Tpost                   & 0.68               & 0.68            & 0.68              & 0.68          \\ \hline
Tthread                 & 0.69               & 0.69            & 0.69              & 0.69          \\ \hline
EF + Gpost              & 0.73               & 0.72            & 0.72              & 0.72          \\ \hline
EF + TfirstPost         & 0.7                & 0.69            & 0.69              & 0.7           \\ \hline
Gpost + TfirstPost      & \textbf{0.8}       & \textbf{0.79}   & \textbf{0.79}     & \textbf{0.79} \\ \hline
Gpost + Tthread         & 0.76               & 0.76            & 0.76              & 0.76          \\ \hline
EF + Gpost + TfirstPost & 0.78               & 0.78            & 0.78              & 0.78          \\ \hline
\end{tabular}
\end{table}

\subsubsection{Summary}
We discovered that the combination of graph and text embeddings had the highest performance. When it comes to text embeddings, we observed that there were only minor differences between the types of text provided to the model, such as a single post or a thread. However, the editors' characteristics did not yield satisfactory results.

\subsection{RQ3 - Do discussions affect the editor's engagement?}

The study aimed to investigate the factors that influence whether a Wikidata editor could be active. 
In this case, the ablation study was focused on editors and the features of editors (EF), text embedding for all their posts (Tavr), and graph embedding for the topology of editors (Geditor) (Table \ref{tab:ML_features}).

Table \ref{tab:results_RQ3} presents the experiment results. 
Similar to the results for predicting a response, the results for predicting the status of editors showed a balance between precision, recall, and F1 scores for the combination of features (e.g., EF + Geditor), suggesting a balanced and high-performing model. In contrast, in the cases of individual features for the characteristics of editors (e.g., EFaccess level), the model tended to make more positive predictions than false positives to all features except for EFage. This suggests that we cannot predict whether an editor is active for sure using only one characteristic. However, we can see that the features for editors, and specifically EFaccess level, presented very low scores. 
These results indicate that editors with higher access levels were not likely to be active compared to basic level users.

Table \ref{tab:results_RQ3} shows that among the features analysed, account age showed the best performance, at $0.72\%$. This may suggest that the longer an editor had been active on Wikidata, the more likely they were to remain active. Upon examining the status of editors with more than six years of account age, we found that $85\%$ of them were active. We further investigated this hypothesis by testing the statistical difference between the account age and the status of editors using the Mann-Whitney U non-parametric test \cite{mcknight2010mann}. The results showed that there was a significant difference between the account age of active and inactive editors, with all the values of account age for the active editors being greater than those of the inactive editors (p-value = 0).

We also discovered that text embedding of editors had an influence on their status, with $0.69\%$ accuracy. This text included all the posts that an editor wrote. 
This result showed that discussions affect the engagement of editors. Additionally, the combination of account age and the content of posts produced the best performance for the status of editors, with $0.75\%$ accuracy.

\begin{table}[h]
\centering
\caption{Performance results of the ablation study for the features individually and in combination. The ML model predicted whether an editor was active.}
\label{tab:results_RQ3}
\begin{tabular}{|l|l|l|l|l|}
\hline
\textbf{Features}   & \textbf{Precision} & \textbf{Recall} & \textbf{Accuracy} & \textbf{F1}   \\ \hline
EFage               & \textbf{0.83}      & \textbf{0.72}   & \textbf{0.72}     & \textbf{0.74} \\ \hline
EFedits             & 0.74               & 0.54            & 0.54              & 0.55          \\ \hline
EFposts             & 0.71               & 0.57            & 0.57              & 0.59          \\ \hline
EFaccess level            & 0.76               & 0.31            & 0.31              & 0.21          \\ \hline
EF                  & 0.83               & 0.73            & 0.73              & 0.75          \\ \hline
Geditor             & 0.63               & 0.63            & 0.63              & 0.63          \\ \hline
Tavr                & \textbf{0.66}      & \textbf{0.69}   & \textbf{0.69}     & \textbf{0.67} \\ \hline
EF + Geditor        & 0.67               & 0.66            & 0.66              & 0.66          \\ \hline
EF + Tavr           & 0.78               & 0.75            & 0.75              & 0.76          \\ \hline
Geditor + Tavr      & 0.65               & 0.65            & 0.65              & 0.65          \\ \hline
EF + Geditor + Tavr & 0.69               & 0.7             & 0.7               & 0.69          \\ \hline
EFage + Tavr        & \textbf{0.78}      & \textbf{0.75}   & \textbf{0.75}     & \textbf{0.76} \\ \hline
\end{tabular}
\end{table}

\subsubsection{Summary}
We found that account age and text embeddings had the highest performance impact on editors' status. Similar to $RQ2$, access level and other characteristics did not show any significant influence.

\section{Discussion}

We analysed the Wikidata discussions by studying the structure of editors' communication networks and exploring what affects the continuity of a discussion and the engagement of editors. Our findings highlight the value of investigating communication interactions to understand what influences engagement and support collaborative communities to improve practice and retain and increase the number of members and their contributions. In this vein, based on our findings, we have developed the following list of recommendations for Wikidata and semantic web communities, which we further elaborate on in this section:
\begin{itemize}
    \item Incorporate instructions and tutorials on how to use talk pages and specialised discussion channels
    \item Develop a post-monitor system similar to HyperNews and GitHub
    \item Develop a mentoring system between long-lasting and new members
    \item Incorporate templates to assist editors in discussing complex KE topics similar to GitHub
\end{itemize}

\textbf{RQ1} 
We found that \textbf{a high number of editors had no edges}, meaning that they did not receive responses to their posts. We showed that the response to a post was related to the status of editors, with $47\%$ of editors who did not receive responses to their posts being inactive. Additionally, our previous study on the content of Wikidata discussions \cite{koutsiana2023analysis} showed that $50\%$ of threads in item talk pages were posts without receiving responses, which could potentially impact the engagement of members, particularly the new ones. To support new members and improve practices, we previously suggested the creation of clear instructions on how to use discussions in Wikidata \cite{koutsiana2023analysis}. Another suggestion to reduce the number of posts with no responses is the development of a system to monitor posts. An annotation system was previously suggested for Wikipedia, similar to other projects like HyperNews \cite{schneider2010content}. This could help classify the type of edits that happen in the content pages. A similar system could be used by Wikidata to classify the posts to \textit{help, question, idea, warning} and so on, and thus bring them to people's attention. This system would allow the community to keep track of posts on talk pages and provide special attention to new members.

We found that Wikidata discussion interactions form a \textbf{small world network}. The small world characteristics: facilitate the propagation of information much faster than random networks; suggests a robust network resilient to node removal, but yet vulnerable to targeted attacks; and benefit scalability, that is, maintaining the small world characteristics even as networks grow in size \cite{leskovec2008planetary,tantardini2019comparing,watts1998collective}. Wikidata could benefit from this to develop strategic planning for tool adoption, or the detection of spam or vandalism. However, the community should be aware of the quick spread of information because this could potentially be harmful in case of deliberate spread of ``fake news'', such as fake practices.

Table \ref{tab:net_met} shows network metrics for four online communities, including Wikidata, forming small world networks. Comparing the network metrics we note that Wikidata is more similar to Wikipedia than to MSN or Slashdot.
A high clustering coefficient indicates a closely connected community where individuals are connected to their friends, and their friends are also connected to each other. It is worth mentioning that the Wikidata network was found to have an \textit{average cluster coefficient} of six and $12$ times higher than the MSN and the Slashdot community, respectively, indicating a stronger sense of cohesion and community, similar to Wikipedia. Additionally, the Wikidata network has the \textit{lowest average} and \textit{longest shortest path} compared to the other three communities, which means that the nodes in the Wikidata network are closer connected to each other, making it easier and faster to traverse from one node to another. 
Furthermore, unlike Slashdot and Wikipedia, Wikidata formed a disassortative network. This means that well-connected editors tended to communicate with low-connected editors within the Wikidata community. This indicates that Wikidata was an inclusive community, more integratively connected than the other two, while the other two had random connections between nodes. This is similar to Wikipedia Wikiprojects, where we find small groups of editors closely collaborating.

\begin{table}[b]
\caption{Network metrics for four online communities.}
\label{tab:net_met}
\begin{tabular}{|p{0.24\linewidth}|p{0.19\linewidth}|p{0.15\linewidth}|p{0.15\linewidth}|p{0.12\linewidth}|}
\hline
\textbf{Community} & \textbf{Average Clustering Coefficient} & \textbf{Average Shortest Path} & \textbf{Longest Shortest Path} & \textbf{Degree Assortativity} \\ \hline
MSN \cite{leskovec2008planetary}                                & 0.1  & 6.6 & 29 & -       \\ \hline
Slashdot (undirected dense network) \cite{gomez2008statistical} & 0.05 & 4   & 9  & -0.04   \\ \hline
Wikipedia (user network) \cite{ingawale2009small}               & 0.9  & 5   & -  & neutral \\ \hline
Wikipedia (mean of Wikiproject networks) \cite{rychwalska2020quality}               & 0.2  & -   & -  & -0.3 \\ \hline
Wikidata                                                                         & 0.6  & 3   & 7  & -0.2    \\ \hline
\end{tabular}
\end{table}

The closely connected community for Wikidata could be explained from the variety of discussion channels, such as \textit{property proposal, administrators' noticeboard}, and \textit{item talk pages}. Some editors might not be limited to one kind of discussion channel, but might rather converse and share opinions to to different channels about different issues, technical or administrative. 
Alternatively, the majority of editors might tend to participate in general issues discussion channels like \textit{request for comments} and subsequently to be connected.
This behaviour may enhance the connection of editors with different responsibilities in the community. Although Wikipedia shares a similar system with various discussion channels, its seems that the two projects have different community dynamics. 
Wikipedia community presents random connections, while Wikidata is an inclusive community with strongly connected groups. 
This may suggest that Wikipedia members may tend to use the article talk pages to communicate more than other discussion channels, while in Wikidata we found a low use of item talk pages \cite{koutsiana2023analysis}

We found that \textbf{$20\%$ of editors made more than 10 posts}, suggesting that they made the most contributions in discussions and they were potentially responsible for key decisions. This is more balanced than the schema.org community \cite{kanza2018does} and the $90-9-1$ rule for participation inequality in social media and online communities \cite{90_9_1}, which suggests that $90\%$ of members are lurkers who never contribute, $9\%$ contribute a little, and $1\%$ account for almost all the activity. This suggests that in Wikidata, there is a higher proportion of editors participating at a low but significant level. In addition, results showed that \textbf{$10\%$ of editors had been members of the community for more than $10$ years}, which is consistent since the earlier stages of the project \cite{sarasua2019evolution}. This indicates that the Wikidata community has a group of committed and experienced members. 
Leveraging their expertise, these members could be advised to guide the development of new systems to improve practices, such as recommendation systems. Furthermore, in the case of creating a system to classify posts to enhance monitoring and limit posts without responses, these experienced members could help by responding to posts and serving as mentors to new members. 
The participation of core contributors in discussions and mentoring of new members has been also emphasised by Hata et al. \cite{hata2022github} for the GitHub projects.

\textbf{RQ2} 
We found that the primary factor influencing whether a discussion received a response was its content. This encompassed the topic and the clarity in defining it.
Our prior work \cite{koutsiana2023analysis} suggested that reasons for no responses in Wikidata could be the numerous items which may leave their talk pages unattended and the lack of guidance for using talk pages compared to Wikipedia. 
In this study \cite{koutsiana2023analysis}, we explored topics of item discussions, showing that $46\%$ of topics were related to KE. The editors discussed content page management, suggested or reported changes, and asked for help and specifications related to Wikidata tasks. The complexity of the subject, particularly for a group of people with diverse KE experience, may make it difficult to express the subjects in the simplest and most informative way. This may mean that not everyone in the community feels that they have the expertise to respond. This argument is enhanced by the findings that the access level of an editor did not have an influence on a post to receive a response. This indicates that even posts from editors with higher editing rights and some experience in KE may remain unanswered.

The complexity of KE topics suggests that guidance in asking questions and discussing changes may increase the possibility of continuing a conversation. This may be possible with the use of templates for specific cases. Wikidata already use templates for specific discussion channels like \textit{request for a comment} and \textit(property proposal) to help the community receive all the information needed to make a decision on a subject. The use of templates for other discussion pages for questions and requests may enhance communication and help new members, in particular, to receive answers. Guidance on discussions has been emphasised previously for the GitHub community to support the participation of members \cite{hata2022github}.

\textbf{RQ3} 
Our findings suggest that the engagement of editors is influenced by their account age and the content of their discussions. We showed that the higher the account age, the more likely an editor was to be active.
This indicates that one important factor for engagement is the time when an editor first joins the community.  
Our previous findings from $RQ2$ showed that the status of editors was related to whether their posts received responses, with $47\%$ of editors who did not receive a response to their posts dropping out of Wikidata, and the rest not attempting to participate in any discussion in the future. These results highlight the influence of no responses on the engagement of editors and emphasise the importance of our previous suggestions to create a system to reduce posts with no responses, especially for item talk pages, which account for $50\%$ of posts did not receive a response \cite{koutsiana2023analysis}. This could improve both the continuation of conversations and the engagement of members in activities, primarily for new members of the community.

In addition, another factor influencing the engagement of editors was the content of their posts.
We previously showed that Wikidata item talk pages had low frequency in conflicts ($7-9\%$ disagreements and up to $1\%$ intensified conflict) \cite{koutsiana2023analysis}.
This could indicate that editors do not receive aggressive or confrontational feedback leading to drop out, but the posts' character, such as the fact that many posts are questions, may have influenced the prediction of editors' status. This is different from open source software where Filippova et al. \cite{filippova2016effects} showed that some types of conflicts impacted the intention of members to email in the project.
As mentioned earlier, in our previous study, we argued that the topics discussed mainly relate to KE activities \cite{koutsiana2023analysis}.
The complexity of KE, especially for members with no prior experience in building KGs, with questions receiving insufficient or confusing feedback, could lead to disappointment and drop outs.
These results highlight again the importance of first experiences related to engagement when new members join the community. The community may need to monitor whether new members ask for help, developing practices for guidance and support to encourage them to stay engaged.
Based on previous findings, we recommended creating an annotation system to monitor posts in order to decrease the number of posts without responses. This system could use advanced technology like generative AI to provide better support for new members, helping them to receive answers and encouraging their continued contributions.

It is worth noting that editor characteristics, such as the access level, did not influence either the communication or the participation of editors. For posts receiving a response, this was expected considering previous findings in $RQ1$, with basic level users being responsible for a greater share of contributions in discussions.


\subsection{Limitations}
For the network analysis, due to the considerable size of the full network, in some cases, we used a smaller random sample to be able to calculate specific metrics. However, the sampling could corrupt the structural properties of the network and result in inconsistent findings. Being able to manage the full size of the network could lead to more robust patterns for metrics like shortest paths and provide extra information with metrics like homophily. Additionally, in our analysis, we assumed that if two editors took part in the same discussion thread, they were connected. However, in threads with a high number of posts, some editors might participate in the same topic of discussion without directly addressing each other. Alternatively, a future study could track the replies to posts in order to connect editors based on who is responding to whom.

For the prediction models, we used a sample of discussion and only quantitative analysis to investigate the impact of text on the course of discussion and the engagement of editors. However, qualitative analysis similar to previous studies about the content discussed \cite{koutsiana2023analysis,kanke2020knowledge} could reveal insights into the influential topics. 
Furthermore, we investigated the influential factors only in item talk pages. In contrast, including different types of discussion channels may provide extra details to the prediction models. Finally, we tested one network structure for the predictions. Experimenting with different designs, such as connecting posts with reply direction, may improve results for the post and editors' topology, giving another understanding of these predictions.




\section{Conclusion}

We investigated the Wikidata editors' interactions through their lens of discussions. We used a mixed method approach using descriptive statistics, network analysis, and graph and text embedding vectors. 
We  found that the Wikidata community forms a small world network similar to MSN, Slashdot, and Wikipedia. However, Wikidata presents stronger connections than the other communities. Our findings suggest that Wikidata is a valuable example for other communities to mimic its practices in order to build collaborative communities that are inclusive and resilient to dropouts. 
Wikidata `popular' editors seem to communicate a lot with less `popular' editors which shows a significant difference between Wikipedia and Wikidata communities. Despite their numerous similarities in terms of the MediaWiki platform, administrative support, and moderation tools, the two communities present differences in communication and connectivity, with Wikidata editors having more connections to each other.
Additionally, we suggested a list of practices to improve communication and engagement with clear instructions about discussions, a post-monitor system, a mentoring system, and communication templates. The recommendation list can help Wikidata and collaborative communities in developing new practices to help new members find their steps in the project to stay engaged, as well as help project designers transmit new tactics to the community.

Beyond Wikidata, our study can provide insights and guidance to semantic web communities on how to build and optimise collaborative ecosystems for sustainability, growth, and quality improvement. Our findings on how Wikidata is connected and what affects communication and participation can be used as a guide not only for general domain KGs and ontologies, but also in specialised domains like healthcare, cultural heritage, and open data initiatives, such as government or public-sector linked data projects.   
Our suggested framework can be used as a base for future studies. Other communities can use this framework and extend it by adding extra specialised features based on the community peculiarities, as well as features on how cultural and linguistic diversity affects communication and participation in global communities. In addition, a future analysis can utilise the framework and the dataset to run the analysis again in order to validate community evolution over time and perhaps after adapting new practices and improvements based on our recommendations. The framework and dataset can be used to deepen the analyses on the topics affecting communication and participation and identify discussions for qualitative analysis or sentiment analysis. A further analysis on the content of the text can justify what makes a post a good candidate to receive an answer and what discourages members from responding to others. In the same vein, it would be interesting to investigate the text that makes editors stop contributing. 
Furthermore, by linking the dataset discussions to the parallel entity activity, a future study could explore how discussions influence productivity. Finally, interviews with Wikidata editors could add valuable insights about their experience in the project, their collaborations, and ideas for improvements.

\bibliography{tgdk-v2021-sample-article}

\begin{thebibliography}{10}

\bibitem{alghamdi2021learning}
Kholoud AlGhamdi, Miaojing Shi, and Elena Simperl.
\newblock Learning to recommend items to wikidata editors.
\newblock In {\em International Semantic Web Conference}, pages 163--181. Springer, 2021.

\bibitem{arazy2011information}
Ofer Arazy, Oded Nov, Raymond Patterson, and Lisa Yeo.
\newblock Information quality in wikipedia: The effects of group composition and task conflict.
\newblock {\em Journal of management information systems}, 27(4):71--98, 2011.

\bibitem{arazy2013stay}
Ofer Arazy, Lisa Yeo, and Oded Nov.
\newblock Stay on the wikipedia task: When task-related disagreements slip into personal and procedural conflicts.
\newblock {\em Journal of the American Society for Information Science and Technology}, 64(8):1634--1648, 2013.

\bibitem{benkler2015peer}
Yochai Benkler, Aaron Shaw, and Benjamin~Mako Hill.
\newblock Peer production: A form of collective intelligence.
\newblock {\em Handbook of collective intelligence}, 175, 2015.

\bibitem{bhattacharya2020impact}
Subhayan Bhattacharya, Sankhamita Sinha, and Sarbani Roy.
\newblock Impact of structural properties on network structure for online social networks.
\newblock {\em Procedia Computer Science}, 167:1200--1209, 2020.

\bibitem{brandes2009network}
Ulrik Brandes, Patrick Kenis, J{\"u}rgen Lerner, and Denise Van~Raaij.
\newblock Network analysis of collaboration structure in wikipedia.
\newblock In {\em Proceedings of the 18th international conference on World wide web}, pages 731--740, 2009.

\bibitem{erdos1959random}
Paul Erdos.
\newblock On random graphs.
\newblock {\em Mathematicae}, 6:290--297, 1959.

\bibitem{falconer2011analysis}
Sean Falconer, Tania Tudorache, and Natalya~F Noy.
\newblock An analysis of collaborative patterns in large-scale ontology development projects.
\newblock In {\em Proceedings of the sixth international conference on Knowledge capture}, pages 25--32, 2011.

\bibitem{fensel2020knowledge}
Dieter Fensel, U~Simsek, Kevin Angele, Elwin Huaman, Elias K{\"a}rle, Oleksandra Panasiuk, Ioan Toma, J{\"u}rgen Umbrich, and Alexander Wahler.
\newblock {\em Knowledge graphs}.
\newblock Springer, 2020.

\bibitem{filippova2016effects}
Anna Filippova and Hichang Cho.
\newblock The effects and antecedents of conflict in free and open source software development.
\newblock In {\em Proceedings of the 19th ACM Conference on Computer-Supported Cooperative Work \& Social Computing}, pages 705--716, 2016.

\bibitem{fisher2017perceived}
David~N Fisher, Matthew~J Silk, and Daniel~W Franks.
\newblock The perceived assortativity of social networks: methodological problems and solutions.
\newblock {\em Trends in Social Network Analysis: Information Propagation, User Behavior Modeling, Forecasting, and Vulnerability Assessment}, pages 1--19, 2017.

\bibitem{gandica2015wikipedia}
Y{\'e}rali Gandica, Jo{\"a}o Carvalho, and F~Sampaio Dos~Aidos.
\newblock Wikipedia editing dynamics.
\newblock {\em Physical Review E}, 91(1):012824, 2015.

\bibitem{garimella2018quantifying}
Kiran Garimella, Gianmarco De~Francisci Morales, Aristides Gionis, and Michael Mathioudakis.
\newblock Quantifying controversy on social media.
\newblock {\em ACM Transactions on Social Computing}, 1(1):1--27, 2018.

\bibitem{gomez2008statistical}
Vicen{\c{c}} G{\'o}mez, Andreas Kaltenbrunner, and Vicente L{\'o}pez.
\newblock Statistical analysis of the social network and discussion threads in slashdot.
\newblock In {\em Proceedings of the 17th international conference on World Wide Web}, pages 645--654, 2008.

\bibitem{hata2022github}
Hideaki Hata, Nicole Novielli, Sebastian Baltes, Raula~Gaikovina Kula, and Christoph Treude.
\newblock Github discussions: An exploratory study of early adoption.
\newblock {\em Empirical Software Engineering}, 27(1):1--32, 2022.

\bibitem{hogan2021knowledge}
Aidan Hogan, Eva Blomqvist, Michael Cochez, Claudia d'Amato, Gerard~de Melo, Claudio Gutierrez, Sabrina Kirrane, Jos{\'e} Emilio~Labra Gayo, Roberto Navigli, Sebastian Neumaier, et~al.
\newblock Knowledge graphs.
\newblock {\em Synthesis Lectures on Data, Semantics, and Knowledge}, 12(2):1--257, 2021.

\bibitem{ingawale2009small}
Myshkin Ingawale, Amitava Dutta, Rahul Roy, and Priya Seetharaman.
\newblock The small worlds of wikipedia: implications for growth, quality and sustainability of collaborative knowledge networks.
\newblock 2009.

\bibitem{joshi2017automatic}
Aditya Joshi, Pushpak Bhattacharyya, and Mark~J Carman.
\newblock Automatic sarcasm detection: A survey.
\newblock {\em ACM Computing Surveys (CSUR)}, 50(5):1--22, 2017.

\bibitem{kanke2020knowledge}
Timothy Kanke.
\newblock Knowledge curation work in wikidata wikiproject discussions.
\newblock {\em Library Hi Tech}, 2020.

\bibitem{kanza2018does}
Samantha Kanza, Alex Stolz, Martin Hepp, and Elena Simperl.
\newblock {What Does an Ontology Engineering Community Look Like? A Systematic Analysis of the schema. org Community}.
\newblock In {\em European Semantic Web Conference}, pages 335--350. Springer, 2018.

\bibitem{kittur2010beyond}
Aniket Kittur and Robert~E Kraut.
\newblock Beyond wikipedia: coordination and conflict in online production groups.
\newblock In {\em Proceedings of the 2010 ACM conference on Computer supported cooperative work}, pages 215--224, 2010.

\bibitem{koutsiana2023analysis}
Elisavet Koutsiana, Gabriel Maia~Rocha Amaral, Neal Reeves, Albert Mero{\~n}o-Pe{\~n}uela, and Elena Simperl.
\newblock An analysis of discussions in collaborative knowledge engineering through the lens of wikidata.
\newblock {\em Journal of Web Semantics}, 78:100799, 2023.

\bibitem{koutsiana2023agreeing}
Elisavet Koutsiana, Tushita Yadav, Nitisha Jain, Albert Meroño-Peñuela, and Elena Simperl.
\newblock Agreeing and disagreeing in collaborative knowledge graph construction: An analysis of wikidata, 2023.
\newblock \href {https://arxiv.org/abs/2306.11766} {\path{arXiv:2306.11766}}.

\bibitem{kumar2017army}
Srijan Kumar, Justin Cheng, Jure Leskovec, and VS~Subrahmanian.
\newblock An army of me: Sockpuppets in online discussion communities.
\newblock In {\em Proceedings of the 26th international conference on world wide web}, pages 857--866, 2017.

\bibitem{leskovec2008planetary}
Jure Leskovec and Eric Horvitz.
\newblock Planetary-scale views on a large instant-messaging network.
\newblock In {\em Proceedings of the 17th international conference on World Wide Web}, pages 915--924, 2008.

\bibitem{lin2015learning}
Yankai Lin, Zhiyuan Liu, Maosong Sun, Yang Liu, and Xuan Zhu.
\newblock Learning entity and relation embeddings for knowledge graph completion.
\newblock In {\em Proceedings of the AAAI conference on artificial intelligence}, volume~29, 2015.

\bibitem{ma2021comprehensive}
Xiaoxiao Ma, Jia Wu, Shan Xue, Jian Yang, Chuan Zhou, Quan~Z Sheng, Hui Xiong, and Leman Akoglu.
\newblock A comprehensive survey on graph anomaly detection with deep learning.
\newblock {\em IEEE Transactions on Knowledge and Data Engineering}, 2021.

\bibitem{massa2011social}
Paolo Massa.
\newblock Social networks of wikipedia.
\newblock In {\em Proceedings of the 22nd ACM conference on Hypertext and hypermedia}, pages 221--230, 2011.

\bibitem{mcknight2010mann}
Patrick~E McKnight and Julius Najab.
\newblock Mann-whitney u test.
\newblock {\em The Corsini encyclopedia of psychology}, pages 1--1, 2010.

\bibitem{moutidis2021community}
Iraklis Moutidis and Hywel~TP Williams.
\newblock Community evolution on stack overflow.
\newblock {\em Plos one}, 16(6):e0253010, 2021.

\bibitem{muller2015peer}
Claudia M{\"u}ller-Birn, Benjamin Karran, Janette Lehmann, and Markus Luczak-R{\"o}sch.
\newblock Peer-production system or collaborative ontology engineering effort: What is wikidata?
\newblock In {\em Proceedings of the 11th International Symposium on Open Collaboration}, pages 1--10, 2015.

\bibitem{newman2003social}
Mark~EJ Newman and Juyong Park.
\newblock Why social networks are different from other types of networks.
\newblock {\em Physical review E}, 68(3):036122, 2003.

\bibitem{patil2023comparative}
Avinash Patil, Kihwan Han, and Sabyasachi Mukhopadhyay.
\newblock A comparative study of {Text Embedding Models} for {Semantic Text Similarity} in {Bug Reports}, 2023.
\newblock \href {https://arxiv.org/abs/2308.09193} {\path{arXiv:2308.09193}}.

\bibitem{piao2021learning}
Guangyuan Piao and Weipeng Huang.
\newblock Learning to predict the departure dynamics of wikidata editors.
\newblock In {\em The Semantic Web--ISWC 2021: 20th International Semantic Web Conference, ISWC 2021, Virtual Event, October 24--28, 2021, Proceedings 20}, pages 39--55. Springer, 2021.

\bibitem{piscopo2018models}
Alessandro Piscopo and Elena Simperl.
\newblock {Who Models the World? Collaborative Ontology Creation and User Roles in Wikidata}.
\newblock {\em Proceedings of the ACM on Human-Computer Interaction}, 2(CSCW):1--18, 2018.

\bibitem{pourhabibi2020fraud}
Tahereh Pourhabibi, Kok-Leong Ong, Booi~H Kam, and Yee~Ling Boo.
\newblock Fraud detection: A systematic literature review of graph-based anomaly detection approaches.
\newblock {\em Decision Support Systems}, 133:113303, 2020.

\bibitem{rychwalska2020quality}
Agnieszka Rychwalska, Szymon Talaga, and Karolina Ziembowicz.
\newblock Quality in peer production systems--impact of assortativity of communication networks on group efficacy.
\newblock 2020.

\bibitem{sarasua2019evolution}
Cristina Sarasua, Alessandro Checco, Gianluca Demartini, Djellel Difallah, Michael Feldman, and Lydia Pintscher.
\newblock The evolution of power and standard wikidata editors: comparing editing behavior over time to predict lifespan and volume of edits.
\newblock {\em Computer Supported Cooperative Work (CSCW)}, 28(5):843--882, 2019.

\bibitem{schneider2010content}
Jodi Schneider, Alexandre Passant, and John~G Breslin.
\newblock {A content analysis: How Wikipedia talk pages are used}.
\newblock In {\em The Web Science Conference 2010 (WebSci '10). Raleigh, North Carolina, USA}, 2010.

\bibitem{simperl2014collaborative}
Elena Simperl and Markus Luczak-R{\"o}sch.
\newblock Collaborative ontology engineering: a survey.
\newblock {\em The Knowledge Engineering Review}, 29(1):101--131, 2014.

\bibitem{solorio2013case}
Thamar Solorio, Ragib Hasan, and Mainul Mizan.
\newblock A case study of sockpuppet detection in wikipedia.
\newblock In {\em Proceedings of the Workshop on Language Analysis in Social Media}, pages 59--68, 2013.

\bibitem{sun2018rotate}
Zhiqing Sun, Zhi-Hong Deng, Jian-Yun Nie, and Jian Tang.
\newblock Rotate: Knowledge graph embedding by relational rotation in complex space.
\newblock In {\em International Conference on Learning Representations}, 2018.

\bibitem{wikimediaContoversies}
Diego Sáez~Trumper and Lydia Pintscher.
\newblock {Research:Identifying Controversial Content in Wikidata}.
\newblock \url{https://meta.wikimedia.org/wiki/Research:Identifying_Controversial_Content_in_Wikidata}, 2021.
\newblock [Online; accessed 10-January-2023].

\bibitem{tantardini2019comparing}
Mattia Tantardini, Francesca Ieva, Lucia Tajoli, and Carlo Piccardi.
\newblock Comparing methods for comparing networks.
\newblock {\em Scientific reports}, 9(1):17557, 2019.

\bibitem{ugander2011anatomy}
Johan Ugander, Brian Karrer, Lars Backstrom, and Cameron Marlow.
\newblock The anatomy of the facebook social graph.
\newblock {\em arXiv preprint arXiv:1111.4503}, 2011.

\bibitem{ugoni1995chi}
Antony Ugoni and Bruce~F Walker.
\newblock The chi square test: an introduction.
\newblock {\em COMSIG review}, 4(3):61, 1995.

\bibitem{vashishth2019composition}
Shikhar Vashishth, Soumya Sanyal, Vikram Nitin, and Partha Talukdar.
\newblock Composition-based multi-relational graph convolutional networks.
\newblock {\em arXiv preprint arXiv:1911.03082}, 2019.

\bibitem{viegas2007talk}
Fernanda~B Viegas, Martin Wattenberg, Jesse Kriss, and Frank Van~Ham.
\newblock {Talk before you type: Coordination in Wikipedia}.
\newblock In {\em 2007 40th Annual Hawaii International Conference on System Sciences (HICSS'07)}, pages 78--78. IEEE, 2007.

\bibitem{vrandecic2013rise}
Denny Vrandecic.
\newblock The rise of wikidata.
\newblock {\em IEEE Intelligent Systems}, 28(4):90--95, 2013.

\bibitem{vrandevcic2014wikidata}
Denny Vrande{\v{c}}i{\'c} and Markus Kr{\"o}tzsch.
\newblock {Wikidata: A Free Collaborative Knowledge Base}.
\newblock {\em Communications of the ACM}, 57(10):78--85, 2014.

\bibitem{wagner2004wiki}
Christian Wagner.
\newblock Wiki: A technology for conversational knowledge management and group collaboration.
\newblock {\em Communications of the association for information systems}, 13(1):19, 2004.

\bibitem{watts1998collective}
Duncan~J Watts and Steven~H Strogatz.
\newblock Collective dynamics of ‘small-world’networks.
\newblock {\em nature}, 393(6684):440--442, 1998.

\bibitem{zangerle2016empirical}
Eva Zangerle, Wolfgang Gassler, Martin Pichl, Stefan Steinhauser, and G{\"u}nther Specht.
\newblock An empirical evaluation of property recommender systems for wikidata and collaborative knowledge bases.
\newblock In {\em Proceedings of the 12th International Symposium on Open Collaboration}, pages 1--8, 2016.

\bibitem{zhou2020universal}
Bin Zhou, Xin Lu, and Petter Holme.
\newblock Universal evolution patterns of degree assortativity in social networks.
\newblock {\em Social Networks}, 63:47--55, 2020.

\bibitem{zhou2020survey}
HJ~Zhou, TT~Shen, XL~Liu, YR~Zhang, Peng Guo, and Jianjun Zhang.
\newblock Survey of knowledge graph approaches and applications.
\newblock {\em Journal on Artificial Intelligence}, 2(2):89--101, 2020.

\end{thebibliography}




\end{document}